\documentclass[prd,amsfonts,onecolumn,superscriptaddress,aps,nofootinbib,10pt]{revtex4}

\pdfoutput=1
\usepackage[utf8]{inputenc}
\usepackage{amsmath,amssymb,mathrsfs}
\usepackage{graphicx}
\usepackage{enumerate}
\usepackage{multirow}
\usepackage{float}
\usepackage[usenames,dvipsnames]{xcolor} 
\usepackage{hyperref}
\usepackage{slashed}
\usepackage{soul}
\hypersetup{colorlinks=true,
		breaklinks=true,
    linkcolor=Blue,          
    citecolor=Plum,        
    urlcolor=YellowOrange           
}

\begin{document}
\title{Jets and Photons Spectroscopy of Higgs-ALP Interactions}

\author{Alexandre Alves}
\email{aalves@unifesp.br}
\affiliation{Departamento de Física, Universidade Federal de São Paulo, Diadema, Brasil}

\author{A. G. Dias}
\email{alex.dias@ufabc.edu.br}
\affiliation{Centro de Ci\^encias Naturais e Humanas, Universidade Federal do ABC, Santo Andr\'e-SP, Brasil}

\author{D. D. Lopes}
\email{lopes.diego@ufabc.edu.br}
\affiliation{Centro de Ci\^encias Naturais e Humanas, Universidade Federal do ABC, Santo Andr\'e-SP, Brasil}

\date{\today}

\begin{abstract}
Axion-like particles (ALPs) and Higgs bosons can interact in scalar sectors beyond the Standard Model, leading the Higgs boson to decay into pairs of gluons and photons through the ALP interaction and giving rise to resonances in the decay products of the process $h\to aa\to gg+\gamma\gamma$, resembling a spectral lines analysis. We explore this signature to constrain an ALP effective field theory formulation and show that the forthcoming runs of the LHC will be capable to probe the ALP-Higgs interaction in the ALP mass range from 0.5 to 60 GeV using an automatized search strategy that adapts to different ALP masses in inclusive jets plus photons final states. Such interaction can also be tested in mass regions where the two and four-photon search channels are currently ineffective.
\end{abstract}
\maketitle
\section{Introduction} 

The scalar sector of the Standard Model (SM) is currently much less constrained than the gauge and matter sectors turning it into a target to the forthcoming runs of the LHC and future colliders as well. Many aspects of the scalar potential remain to be elucidated including the possibility of new interactions. 
One plausible possibility is that the Higgs boson, $h$, has some sort of coupling to a light pseudoscalar, $a$, allowing the decay $h\rightarrow aa$. Such pseudoscalars may arise as pseudo Nambu-Goldstone bosons of spontaneously broken quasi exact-symmetries, potentially involved in long sought solutions of problems left open by the SM. Well known examples are axions and axion-like particles ALPs (for reviews of axions and ALPs see, for example, \cite{Zyla:2020zbs,Kim:2008hd,Jaeckel:2010ni,Ringwald:2012hr,Marsh:2015xka,DiLuzio:2020wdo,Agrawal:2021dbo,Bauer:2018uxu}). Axions are the pseudo Nambu-Goldstone bosons of a peculiar  anomalous global chiral U(1)$_{PQ}$ symmetry, which is required to solve the strong CP problem through the Peccei-Quinn  mechanism~\cite{Peccei:1977hh,Peccei:1977ur,Weinberg:1977ma,Wilczek:1977pj}. The axion mass and its couplings to gauge and matter fields are inversely proportional to the energy scale in which the U(1)$_{PQ}$ symmetry is broken spontaneously. In most common viable axion models, that scale is very high leading to tiny masses and couplings which, although being aimed by several experiments and confronted with astrophysical data \cite{Zyla:2020zbs,DiLuzio:2020wdo,Agrawal:2021dbo}, cannot produce direct observable effects in collisions at the LHC~\footnote{It should mentioned that some constructions actually predict axions that could be observed at colliders \cite{Rubakov:1997vp,Dimopoulos:2016lvn,Hook:2014cda}.}. On the other hand, ALPs do not necessarily share a mass-coupling relation like the axion so their masses and couplings to SM particles become independent parameters and more amenable for phenomenological studies at the same time it gets easier to avoid experimental constraints.
 
 Many studies have already addressed an augmented Higgs scalar sector with pseudoscalars in ultra-violet completed theories. Good examples are the two Higgs doublet models \cite{Branco:2011iw}, with the pseudo-scalar mass arising from a soft broken global symmetry in the scalar potential, and the Next-to-Minimal Supersymmetric Standard Model~\cite{Miller:2003ay,Martin:1997ns}. However, instead of a top-bottom approach, given the many situations where we expect an extended scalar sector, we might think of looking firstly for effective interactions between the SM and the new scalar particle~\cite{Brivio:2017ije,Bauer:2017ris,Bauer:2020jbp}. We will then focus on generic ALPs using the effective theory approach, i.e. without assuming a particular model for them, looking for the their masses and couplings to the Higgs boson, gluons and photons that could give a potential signal at the LHC, within a region of the parameter space not previously explored.

 Under the assumption that the ALP is linked to an approximate global U(1) symmetry, which is realized non-linearly as the shift symmetry $a\to a+\alpha$, with $\alpha$ constant,  the lower order interaction occurs through a dim-6 operator of the form  $\frac{1}{\Lambda^2}(\partial_\mu a)(\partial^\mu a)\phi^\dagger \phi$, where $\Lambda$ is related to the U(1) symmetry breaking scale occurring above the electroweak scale and $\phi$ is the Higgs doublet. After electroweak symmetry breaking, this operator induces the Higgs to ALP decays if it is kinemactically allowed. Effective operators of ALP-gluons and ALP-photons are generally expected to be generated at one loop level in ultra-violet completed theory where its couplings with new heavy charged and colored fermions $\psi$ occur through renormalizable operators like $iy\,a\,\overline{\psi}\gamma_5\psi$.
 In the case of ALPs, besides their probable role in the solutions to unanswered questions in particle physics, light axion/ALP extend their scope to other observed anomalies if we consider, for example, couplings to photons and leptons that might explain the $4.2\sigma$ discrepancy of the $(g-2)_\mu$ data~\cite{Bauer:2017nlg}, the excess in excited Beryllium decays~\cite{Ellwanger:2016wfe} and the excess in the gamma rays flux from the center of the galaxy observed by Fermi-LAT~\cite{doi:10.1146/annurev-nucl-101916-123029, Berlin:2014tja}, for instance.

  At the LHC, theoretical studies and experimental searches for the SM Higgs boson decaying to ALPs have involved final states with jets, leptons, $Z$ bosons and photons in various ALP mass regions~\cite{Aad:2020hzm,Aaboud:2018iil,Aad:2020rtv,Sirunyan:2019gou,Sirunyan:2020eum,Aad:2015bua,Aaboud:2018gmx,Bauer:2017nlg,Bauer:2017ris,dEnterria:2021ljz}.
  If the ALP mass is below the lightest pion threshold, the decays to photons become the most relevant unless the coupling to electrons and muons get strong. Above the pion threshold, however, the decays to gluons involve detecting rare signals in an overwhelming background of QCD jets. The preferred option thus remains to search for multiphoton signals but with rates that die out as the ALP decays to hadrons become dominant.
 
 Assuming that couplings to gluons and photons are dominant, detecting the Higgs decays to ALP pairs, $h\to aa$, can be done in the following basic channels depending on the ALP mass, $m_a$: (1) $aa\to \gamma\gamma\gamma\gamma$, (2)  $aa\to gg\gamma\gamma$, (3) $aa\to gggg$. If the ALP is light enough, its decay yields get colimated in the lab frame and hit a single cell of the detector giving rise to signals like: (4) $aa\to \gamma\gamma$, (2)  $aa\to j\gamma$, (3) $aa\to jj$. However, due the detector finite resolution, any signal  of the form $nj+m\gamma$, with $0\leq (n,m) \leq 4$, $2\leq n+m\leq 4$, is expected. Hereafter, $j$ denotes a gluon jet as we are assuming that the ALP couples predominantly to gauge bosons.
 
 Whatever is the ALP mass, the decays to jets, exclusively, cannot be disentangled from QCD backgrounds leaving the multiphotons as the prime option. Indeed, if we do not consider ALP-fermion couplings as relevant, $h\to aa\to \gamma\gamma$ is the best channel to constrain the Higgs-ALP coupling if $m_a\lesssim 400$ MeV, and $h\to aa\to \gamma\gamma\gamma\gamma$ is a good option for $m_a>$ few GeV. However, the mass range from $400\; \hbox{MeV} \lesssim m_a \lesssim$ few GeV, where the hadrons decays are turned on remains not accessible for multiphoton and multijet states.    
 
 In this work, we show that this mass gap can be probed at the 13 TeV LHC in the $pp\to h\to aa\to j(j)+\gamma(\gamma)$ channel where the ALP decays get eventually colimated enough to form a single photon or jet hit in the detector. In the SM, such Higgs decays are either forbidden by color charge conservation or loop-suppressed and observing such a signal would be an striking evidence for Higgs--ALP interactions. Our inclusive signal considers at least one isolated jet and one isolated photon, that is it, we accept events with one and two isolated photons plus any number of isolated jets,
 what leads us to tackle many background sources with jets and photons. In order to clear our signals from the backgrounds, we employ an automatized kinematic cuts strategy which permits us to probe the effective Higgs-ALP interaction assuming various model scenarios in the mass range from 0.5 GeV, where the hadronic decays open, up to 60 GeV, the Higgs to ALP decay threshold. In special, our analysis is effective in the mass gap from 0.5 to $\sim 10$ GeV where two and four photons final state searches lose sensitivity.  Our search strategy explores the Higgs and ALP resonances in photons and jets final states by identifying the various spectral lines in the invariant mass distributions in order to determine whether the Higgs might contain ALPs in its composition. 
 
 The $h\to aa\to gg\gamma\gamma$ channel has already been investigated in Refs.~\cite{Aaboud:2018gmx,Martin:2007dx}. The ATLAS Collaboration has placed limits of the Higgs production cross section times the branching ratio of $h\to aa\to gg\gamma\gamma$ in the 20--60 GeV mass region. We take these limits into account in our analysis. In Ref.~\cite{Martin:2007dx}, which appeared before the Higgs boson discovery, the ALP pair was assumed as being the dominant Higgs decay mode for Higgs masses from 80 to 160 GeV. Contrary to ours, those analyses consider just the fully resolved final state with two isolated jets and two isolated photons with a fixed cut strategy. As we are going to show, to explore the whole mass region from 0.5 to 60 GeV with maximum sensitivity, it is necessary to focus on different resonances and adjust the selection requirements for each ALP mass.
 
 The paper is organized as follows: in Section~\ref{sec:eft}, we present the effective interaction Lagrangian and the relevant decay widths of the Higgs and the ALP bosons; in Section \ref{sec:constraints}, we enumerate and discuss the collider constraints imposed on the search for the ALP signals at the LHC; Sections \ref{sec:simulation} and \ref{sec:models} are devoted to details of simulations, analysis and scanning of the parameters model; we present our results in Section \ref{sec:results}; Section~\ref{sec:conclusion} contains the conclusions.
 

\section{Effective Lagrangian and decay widths}
\label{sec:eft}

Let us consider the following effective interaction Lagrangian
\begin{equation}
\begin{split}
\label{eq:Lagrangian2}
\mathscr{L}_{eff} &= \frac{k_{BB}}{\Lambda} a B^{\mu \nu} \tilde{B}_{\mu \nu} + \frac{k_{WW}}{\Lambda} a W_i^{\mu \nu} \tilde{W}_{\mu \nu}^i + \frac{k_{GG}}{\Lambda} a G_a^{\mu \nu} \tilde{G}_{\mu \nu}^a \\ 
& + \frac{\sqrt{2}vC^{eff}_{ha}}{\Lambda^2} (\partial_{\mu} a) (\partial^{\mu} a) h + \frac{C^{eff}_{ha}}{2\Lambda^2} (\partial_{\mu} a) (\partial^{\mu} a) h h, 
\end{split}
\end{equation}
where $h$ is the Higgs field, and $W_{\mu\nu}$, $B_{\mu\nu}$, and $G^a_{\mu\nu}$ are the $SU(2)_L$, $U(1)_Y$ and $SU(3)_C$ field strength tensors, respectively. The pseudoscalar ALP field is denoted by $a$. The couplings of the ALP with the weak gauge interaction eigenstates $B$, and $W^i,i=1,2,3$ are given by $k_{BB}$ and $k_{WW}$, respectively, while $k_{GG}$ is the coupling between ALPs and gluons.  The dual field $\tilde{B}_{\mu\nu}$ is defined as $\frac{1}{2}\varepsilon_{\alpha\beta\mu\nu}B^{\alpha\beta}$, with the totally anti-symmetric tensor being such that $\varepsilon_{0123}= 1$, and analogously to the other fields. The last two terms represent the triple and quartic ALP-Higgs interactions, $v$ is the SM vacuum expectation value, and $\Lambda$ is the BSM scale and respect a shift symmetry $a\to a+\alpha$ of the ALP field. 

The relevant decay widths for $h \to a a$, $a \to \gamma \gamma$ and $a \to g g$ due to the Lagrangian \eqref{eq:Lagrangian2} are given by
\begin{eqnarray}
&& \Gamma (h \to a a) = \frac{v^2 |C^{eff}_{ha}|^2}{32 \pi \Lambda^4} m_h^3 \left(1 - \frac{2 m_a^2}{m_h^2}\right)^2 \sqrt{1 - \frac{4 m_{a}^2}{m_{h}^2}},\nonumber \\
&& \Gamma (a \to \gamma \gamma)= \frac{(k_{BB} c_W^2 + k_{WW} s_W^2)^2}{4 \pi \Lambda^2} m_{a}^3,\nonumber \\
&& \Gamma (a \to g g) = \frac{8 k_{GG}^2}{4 \pi \Lambda^2} m_{a}^3\; ,
\label{eq:widths}
\end{eqnarray}
and the total width of the ALP is computed from
    \begin{eqnarray}
    \Gamma_a &=& \Gamma(a \to \gamma \gamma) +  \Gamma(a \to g g),
    \label{eq:w_total}
    \end{eqnarray}
assuming that the ALP couplings to gauge bosons are dominant. The ALP mass range that we are interested in does not allow decays to $W$ and $Z$ bosons.

 In principle, the model can be parametrized by 5 independent parameters: the ALP mass, $m_a$, and the four effective couplings $k_{BB}/\Lambda$, $k_{WW}/\Lambda$, $k_{GG}/\Lambda$, and $C_{ah}^{eff}/\Lambda^2$. The coupling of the ALP to photon, however, occurs only via the combination $k_{\gamma\gamma}/\Lambda\equiv (k_{BB} c_W^2 + k_{WW} s_W^2)/\Lambda$, with the sine(cossine) of the electroweak mixing angle defined as $s_W(c_W)\equiv \sin\,\theta_W(\cos\,\theta_W)$. Moreover, once we are considering only a minimum setup where ALPs decay only to gluons and photons, their branching ratios must add to unity. This way, the signal can be conveniently parametrized by the ALP mass, $m_a$, the ALP to photons branching ratio, $BR(a\to \gamma\gamma)$, and the effective Higgs-ALP coupling, $C_{ah}^{eff}/\Lambda^2$. Next, we present the collider constraints taken into account in our analysis.

\section{Collider Constraints}
\label{sec:constraints}

 The $h\to aa\to j(j)+\gamma(\gamma)$ channel can be probed at the LHC for those model parameters that lead to short lived ALPs with masses above the hadronic threshold. We also must avoid parameters already excluded by previous searches in the Higgs channels and searches dedicated exclusively to ALP detection. Right below, we present the 95\% CL constraints imposed on the points of the parameters space that can be probed at the LHC. We assume that the Higgs boson is produced in gluon-fusion with the SM rate.
    \begin{enumerate}
    \item{\underline{ALP lifetime}} -- the ALP must decay before the electromagnetic calorimeter. The distance that the ALP  produced in the decay of the Higgs boson travels from the interaction point before decaying is given approximately by 
    \begin{eqnarray}
    l_{decay} \approx \frac{m_{h}}{m_a \Gamma_a} \times 10^{-16}\; {\rm m}, 
    \label{eq:ldecay}
    \label{ldecay}
    \end{eqnarray}
    where the ALP total width is given by Eq.~\eqref{eq:w_total}.
    
    This way, in order to have a decay inside the ATLAS or the CMS detector, $l_{decay}\sim 1$ meter, we require that
    \begin{equation}
    \frac{m_{a}}{m_h} \Gamma_a > 10^{-16}\; {\rm m}. 
    \label{eq:ldecay2}
    \end{equation} 
    
    \item{\underline{$h \to jj\gamma\gamma$}} -- the direct search in the channel of our interest from the ATLAS Collaboration~\cite{Aaboud:2018gmx}. In this work, limits from 
    \begin{equation}
    \frac{\sigma_H}{\sigma_{SM}}\times BR(h\to aa\to gg\gamma\gamma) < BR_{95}^{jj\gamma\gamma}(m_a),
    \label{eq:hjjaa}
    \end{equation}
    in the 20--60 GeV ALP mass region, were obtained. Here, $\sigma_H$ and $\sigma_{SM}$ are the inclusive production cross section of $pp\to h$ in the BSM and the SM scenarios, respectively,
    and $BR_{95}^{jj\gamma\gamma}(m_a)$ are the ALP mass dependent limits on that ratio. These limits are not so strong as the next ones that we considering, but they are important once they target part of the search channel that we are studying.
    

    \item{\underline{$h \to \gamma\gamma\gamma\gamma$}} -- 
    the ATLAS Collaboration has performed a search for $h\to aa\to \gamma\gamma\gamma\gamma$ in the mass range of 10 to 62.5 GeV in Ref.~\cite{Aad:2015bua} resulting in the following limit
    \begin{equation}
        BR(h\to aa)\times BR^2(a\to\gamma\gamma) < (3\textendash4)\times 10^{-4}.
        \label{eq:4gamas_high}
    \end{equation}
    
     \item{\underline{The Higgs boson total width}} -- the Higgs boson total width is constrained, from on-shell and off-shell Higgs production in the four-lepton final state~\cite{Sirunyan:2019twz}, to be: $\Gamma_h = 3.2^{+2.8}_{-2.2}$ MeV, while the SM theoretical value is $\Gamma_h = 4.1^{+5.0}_{-4.0}$ MeV. Thus, if we suppose that the difference between the upper bound of $\Gamma^{exp}_h$ and the SM value of 4.07 MeV is due the contribution of a single new channel, then 
    \begin{equation}
    BR(h \to a a) \le \frac{6.0 - 4.1}{6.0} = 32.2\%\; .
    \label{eq:h_total}
    \end{equation}
    
    We will assume that dimension-5 operators like $\frac{C_{zh}^{eff}}{\Lambda}(\partial_\mu a)Z^\mu h$
    do not play a role in our analysis, that is it, we assume $C_{ah}^{eff} >> C_{zh}^{eff}$. Our analysis is complementary to those carried out for dimension-5 operators from Refs.~\cite{Bauer:2017ris,Brivio:2017ije}. 
    
    \item{\underline{Photonic decays of the $Z$ boson}} -- the ALP masses considered in this work permits $Z\to a\gamma$ decays~\cite{Jaeckel:2015jla,Alves:2016koo}. In special, the limits imposed on $Z\to\gamma\gamma(\gamma)$ from the CDF~\cite{Aaltonen:2013mfa} and the ATLAS~\cite{Aad:2015bua} Collaborations should be taken into account for ALP masses in the 0.5 to 60 GeV region. Taking the strongest limit from the ATLAS search for $Z$ bosons decaying into three photons~\cite{Aad:2015bua}, we have
    \begin{equation}
        \frac{m_Z^3}{6\pi\Gamma_Z}\left(1-\frac{m_a^2}{m_Z^2}\right)^3 s_W^2 c_W^2\left(\frac{k_{BB}}{\Lambda}-\frac{k_{WW}}{\Lambda}\right)^2\times BR(a\to\gamma\gamma) < 2.2\times 10^{-6}, 
        \label{eq:Zphoton}
    \end{equation}
    where $m_Z=91.18$ GeV and $\Gamma_Z=2.5$ GeV are the $Z$ mass and width, respectively. For $m_a\in [0.5,60]$ GeV and $BR(a\to\gamma\gamma)$ from 1\% to 50\%, this constraint implies $\left|\frac{k_{BB}}{\Lambda}-\frac{k_{WW}}{\Lambda}\right|\lesssim 10^{-4}\; \hbox{to}\; 10^{-5}$ that can be achieved for $k_{BB},k_{WW}\lesssim 10^{-1}\; \hbox{to}\; 10^{-2}$ and $\Lambda\gtrsim 1$ TeV. 
    
    \end{enumerate}
    
 The ATLAS Collaboration put limits on Higgs-ALP couplings from searches for $h\to\gamma\gamma$ for the ALP mass range of 100--400 MeV~\cite{ATLAS-CONF-2012-079}. As we are going to discuss, we estimate, and also observe in the simulations, that resonances in two resolved photons can indeed be observed for $m_a\gtrsim 10$ GeV, and the 0.5--10 GeV mass region is a transition zone where photons from ALP decays begin to be resolved in the detectors. Therefore, we do not place Higgs to two photons constraints on the model once it is difficult to ascertain the validity of those constraints in this transition mass range.
 
 Collider searches in diphotons and dijets look for high mass resonances and also do not constraint the ALPs that we are considering in this work. Astrophysical and other low-energy experiments which hunt axions and axion-like particles have sensitivity to ALPs which are much lighter than the ones that we are focusing, in the 0.5 to 60 GeV mass region. For a good account of constraints on masses and couplings of effective ALP models, see Ref.~\cite{Bauer:2017ris}.  

 In the Section~\ref{sec:models}, we show the relative number of parameters space points that survive the constrains impose by Eqs.~\eqref{eq:ldecay}--\eqref{eq:Zphoton}. Now, we give details about the simulations and the search analysis.

\section{Simulation details and search analysis}
\label{sec:simulation}
 
We simulated events for the signal $gg \to h \to aa \to gg \gamma\gamma$ leading to at least one isolated jet and one isolated photon, that is it, one or two isolated photons plus any number of isolated jets, and its main SM backgrounds: the prompt production $j\gamma$ with one additional jet, $jj\gamma$; the double photon production with one and two additional jets $j\gamma\gamma$, and $jj\gamma\gamma$; the dijet production $jj$, where the decay $\pi^0\to\gamma\gamma$ and other scalar mesons inside the jets produce one or two hard isolated photons. The Higgs decay to gluon jets and photons is either forbidden by color charge conservation or suppressed and can be safely ignored. 

The simulations were performed at leading order for the 13 TeV LHC using \texttt{MadGraph5}~\citep{Alwall:2014hca}. Parton hadronization and showering were simulated with \texttt{Pythia8}~\citep{Sjostrand:2007gs} while detector effects were taken into account from \texttt{Delphes3}~\citep{deFavereau:2013fsa}. 

A first difficulty to overcome is the huge QCD $jj$ background which amounts to $10^{11}$ femtobarns at the 13 TeV LHC. A hadronic jet contains pions and other mesons that might decay into hard and isolated photons that mimic the signal. For commonly adopted photon isolated criteria, one in every $10^4$ jets, approximately, will look like an isolated photon which means that we should expect around $10^7$ fb cross section of fake $j+\gamma$ from the $jj$ background, that turns out to be similar to the irreducible prompt photon background cross section. In order to reduce this potentially dangerous background source we demand a tight photon isolation criteria inspired in Refs.~\cite{Aaboud:2017cbm,Aad:2015bua}. A photon is considered to be isolated if particles with $p_T>0.5$ GeV inside a cone of $\Delta R=0.4$ centered around the photon 3-momentum do not deposit a total fraction of their transverse momentum above 0.05 of the photon $p_T$. Jets are reconstructed using the anti-kt algorithm with a radius parameter of $R = 0.4$ with a minimum transverse momentum of 20 GeV. The tight isolation criterion guarantees that a small fraction of multijet events will produce an isolated photon making it easier to suppress the huge $jj$ background~\cite{Aaboud:2017cbm}. Our simulations show that these tight isolation criteria reduce the $jj$ background to less than 10\% of the irreducible $j+\gamma$ background\footnote{We also checked that our simulations reproduce very well the $j+\gamma$ cross section obtained in Ref.~\cite{Aaboud:2017cbm} with their isolation and selection criteria.}. Nevertheless we analyse the possible effect of remaining $jj$ samples as a systematic uncertainty effect in the main background estimates. This isolation requirement comes at the cost of also suppressing the signals of ALPs with masses at the transition mass region where not too colimated ALP decays start to get resolved by the detectors. 

In the LHC detectors, reconstructed photons are obtained from clusters of energy deposited in an electromagnetic calorimeter (ECAL) of finite granularity. When two particles hit an small area of the detectors, the signal might be interpreted as a single hit. The distance in the $\eta\times \phi$ plane between the particles from the ALP decay which comes from a Higgs boson decay is given by $\Delta R \sim \frac{2 m_a}{p_T} \leq \sqrt{(\Delta\phi)^2+(\Delta\eta)^2}$, where $p_T$ denotes the transverse momentum of the ALP, and the azimuth and rapidity resolutions of the LHC detectors are $\Delta\phi$ and $\Delta\eta$, respectively. Thus ALP masses up to $\sqrt{(\Delta\phi)^2+(\Delta\eta)^2}\times p_T/2$ GeV lead to unresolved decay products, two photons form a single "photon-jet" and two gluons form a single hadronic jet. Adopting the default resolutions of \texttt{Delphes3} in the rapidity region of $|\eta|<3.0$, $\Delta\phi=\Delta\eta=0.02$, and given that the bulk of events have $p_T\sim m_h/2$ ($m_h=125$ GeV is the Higgs boson mass), we should expect that most of the signal events are still colimated for ALP masses up to $\sim 1$ GeV. As we discussed in Section~\ref{sec:constraints}, the ATLAS Collaboration put limits on ALP signals for $m_a < 0.4$ GeV from their search in $h\to\gamma\gamma$ with colimated ALP decays. Anyways, the ALP mass region of a few GeV is a transition region between the resolved and unresolved jets and photons signals. Collider searches for ALP signals into resolved and unresolved photon-jets have been studied, for example, in Refs.~\cite{Ellis:2012zp,Allanach:2017qbs,Sheff:2020jyw,Wang:2021uyb}.

Because we are interested in that mass region where we expect to observe resolved and unresolved photons and jets, we adopted an inclusive search channel accepting at least one isolated jet and one isolated photon with
\begin{eqnarray}
&&    p_T^j > 30\;\hbox{GeV},\; |\eta_j| < 2.4, \nonumber\\
&&    p_T^\gamma > 20\;\hbox{GeV},\; |\eta_\gamma| < 2.4, \nonumber\\
&&    \Delta R_{j\gamma}>0.4 \; .
\label{eq:basic_cuts}
\end{eqnarray}

 As we said, the backgrounds simulations included the prompt photon+jet production, $q\bar{q}\to g+\gamma$ and $gq\to q+\gamma$ taking into account all the QCD and QED hard contributions up to an extra jet and an extra photon. We can classify these background sources in four classes: (1) $j\gamma$, (2) $jj\gamma$, (3) $j\gamma\gamma$, and (4) $jj\gamma\gamma$. We found that contributions from $Z$, $W$ and Higgs decays are negligible after cuts. ALP pair production and associated ALP plus photons and jets~\cite{Ebadi:2019gij} could also mimic the signal from Higgs decays to ALPs but, as we are going to discuss, the photons and jets are required to reconstruct the Higgs mass peak as close as possible what effectively suppresses those contributions to the signal. Now, we discuss the selection criteria to detect the signals.

\subsection{Selection criteria -- Higgs-ALP spectroscopy}

\begin{figure}[t!]
\includegraphics[scale=0.45]{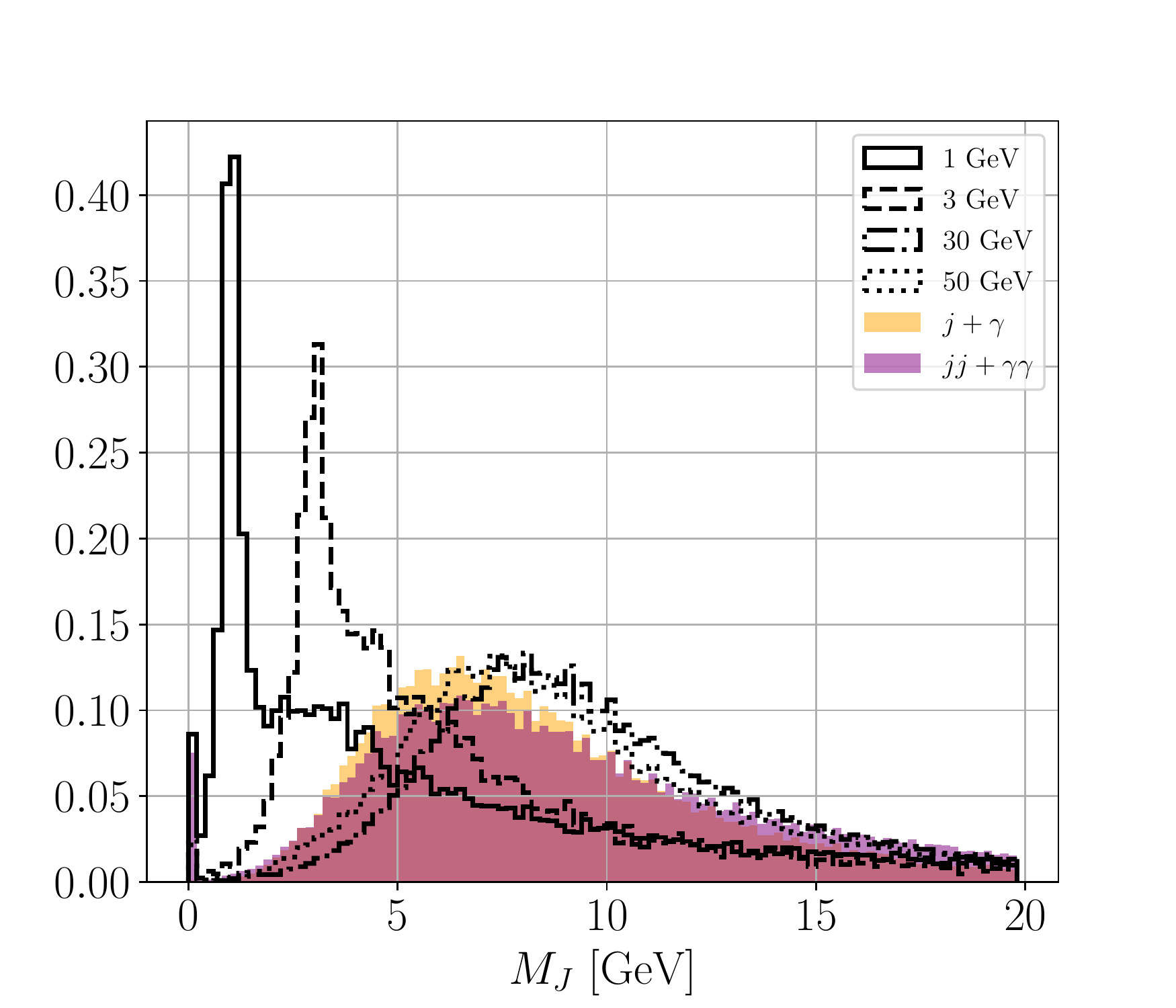}
\includegraphics[scale=0.45]{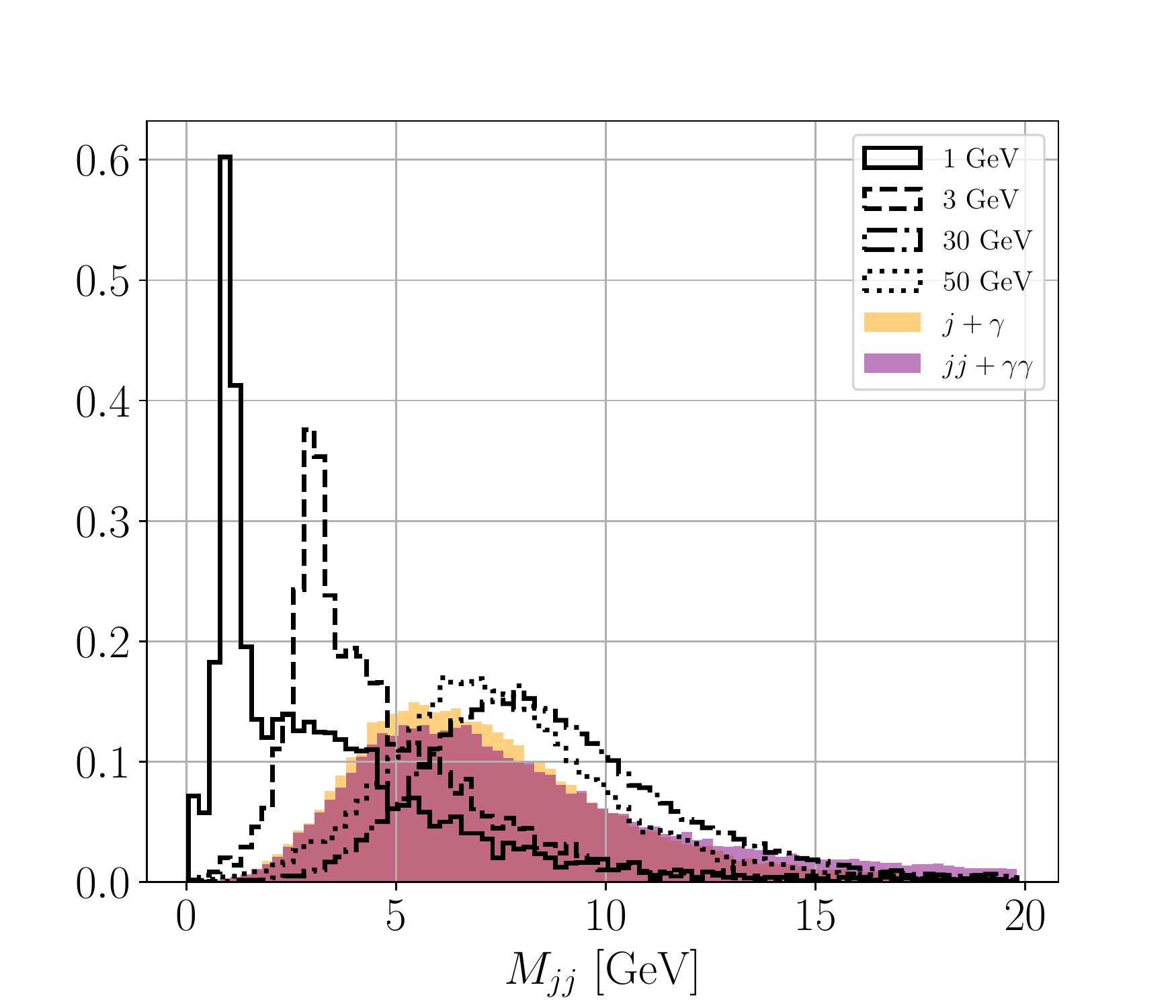}\\
\includegraphics[scale=0.45]{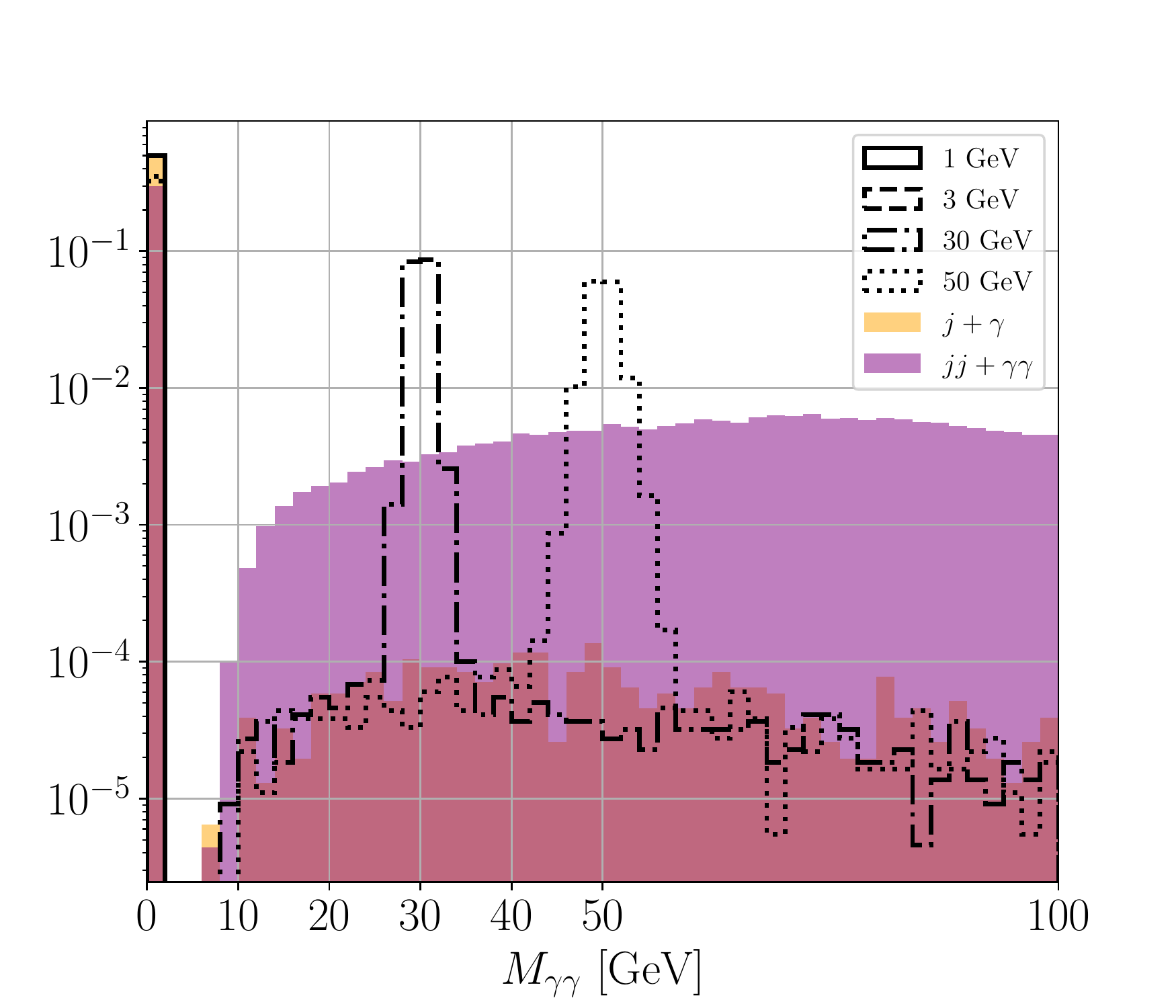}
\includegraphics[scale=0.45]{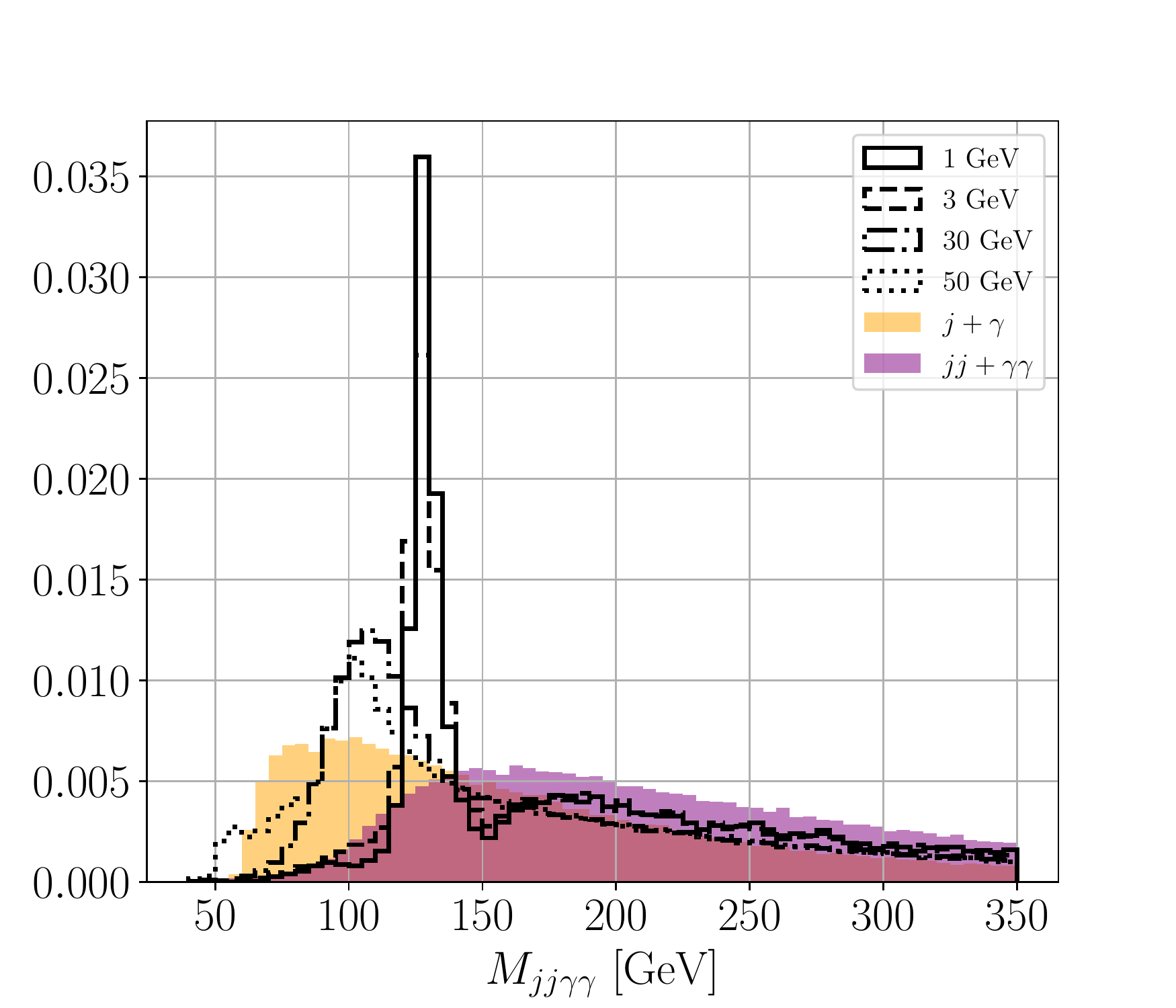}
\caption{The normalized invariant mass distributions of photons pairs (lower left), jets pairs (upper right), the mass of the leading $p_T$ jet, $J$ (upper left), and the mass of $j(j)+\gamma(\gamma)$ pairs (lower right). We display the distributions for ALP masses of 1, 3, 30 and 50 GeV, and for the $j+\gamma$ and $jj\gamma\gamma$ backgrounds.}
\label{fig:lines}
\end{figure}

 The most discerning features of the signals are their invariant mass spectra, which could be think of distinct spectral lines. In Fig.~\eqref{fig:lines}, we display the invariant mass distribution of $jj$ pairs and the leading $p_T$ jet ($J$) mass at the upper row, and the $\gamma\gamma$ and the $j(j)+\gamma(\gamma)$ masses at the lower row. 

The $M_{jj\gamma\gamma}$ shows the jets plus photons invariant mass. This variable is computed with up to two leading photons and two leading jets, comprising also events with just one jet and one photon solely. It is clear from the lower panels of Fig.~\eqref{fig:lines} that a clear peak at the Higgs mass is visible at $M_{jj\gamma\gamma}$ for small ALP masses, up to 10 GeV actually. For higher masses,  the $j(j)+\gamma(\gamma)$ mass does not peak right at the Higgs mass anymore. This is due the tight photon isolation criteria needed to suppress the multijet QCD background which makes the identification of two isolated photons rarer than only one photon. As a consequence, for higher ALP masses where two resolved photons and jets were expected, many events turn out to have just one tightly isolated photon and the combined masses of the final state particles do not add to the Higgs mass. On the other hand, for lower masses, the identified leading jet and photon come from colimated pairs and reconstruct the ALPs and the Higgs, consequently. 

In those cases where two isolated photons can be identified in the event, we observe a peak in the $\gamma\gamma$ mass, as we see in the lower left panel of Fig.~\eqref{fig:lines}. However, if the event contains just one isolated photon, even if it is a photon-jet composed of two colimated photons, it is interpreted as a single photon of null mass. That is why we see a populated been at 0 GeV in the $M_{\gamma\gamma}$ distribution. 

In events where just one isolated photon is detected, we switch to one or two jets masses to observe the ALP resonance, as we can see in the upper panels of Fig.~\eqref{fig:lines}. In those events where just one isolated jet can be identified, we observe that the jet mass exhibits a peak for lighter masses (see the upper left panel of Fig.~\eqref{fig:lines}), once the reconstructed jet will contain the two jets of the event colimated into a single fat jet. This explains the similarity between the $M_J$ distribution explained and $M_{jj}$, which is the mass of up to two leading jets, including the events with just one identified jet. In practice, we could use just $M_{jj}$ to look for light ALP peaks in cases where at least one isolated jet is present, but we also used $M_J$ to place cuts. The difficulty arisen from the tight photon isolation criteria in the task to identify the Higgs and the ALP resonances calls for an automatic and optimized way to identify the resonances and the signal rich regions of the features space in the search of ALPs with unknown masses.

\begin{figure}[t!]
\includegraphics[scale=0.45]{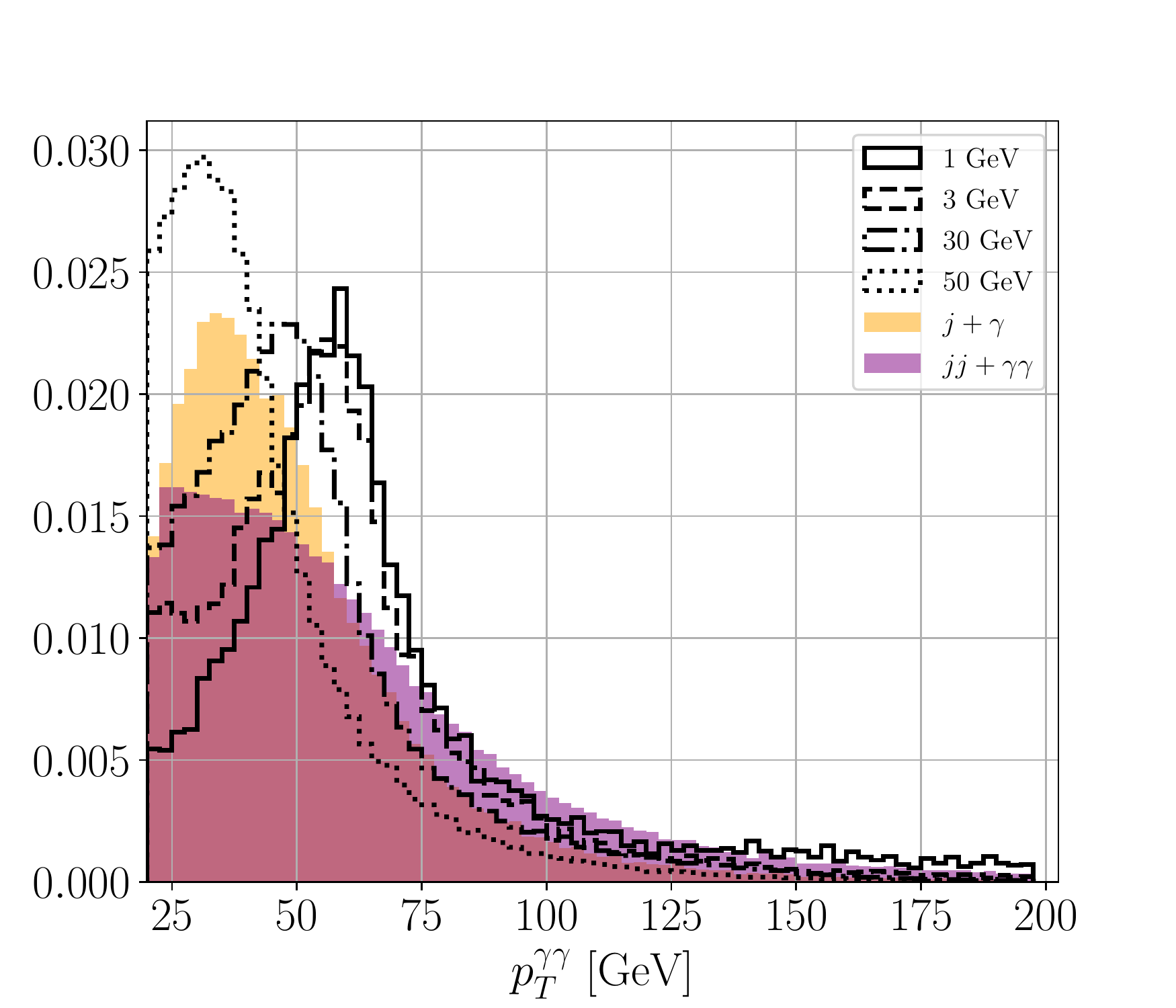}
\includegraphics[scale=0.45]{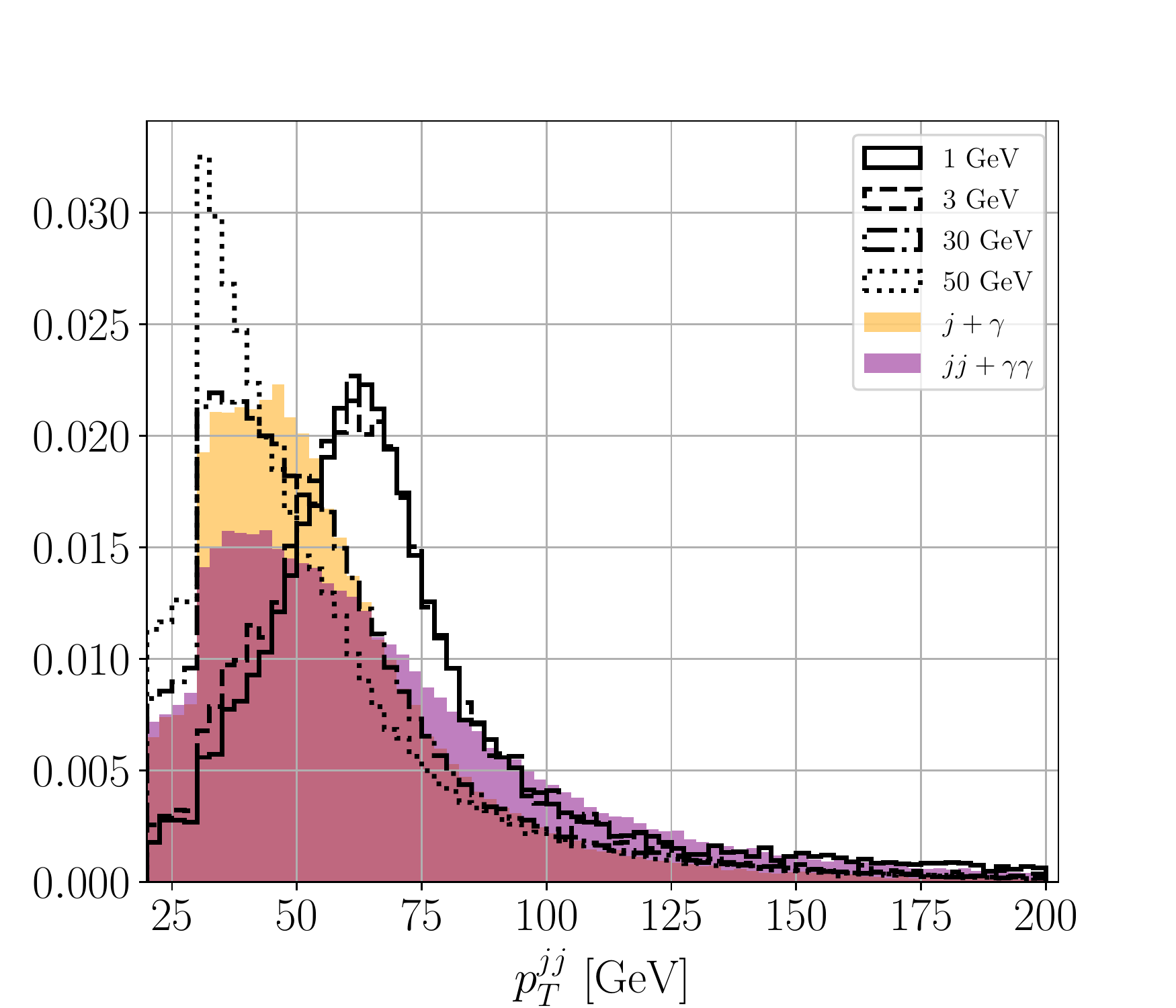}\\
\includegraphics[scale=0.45]{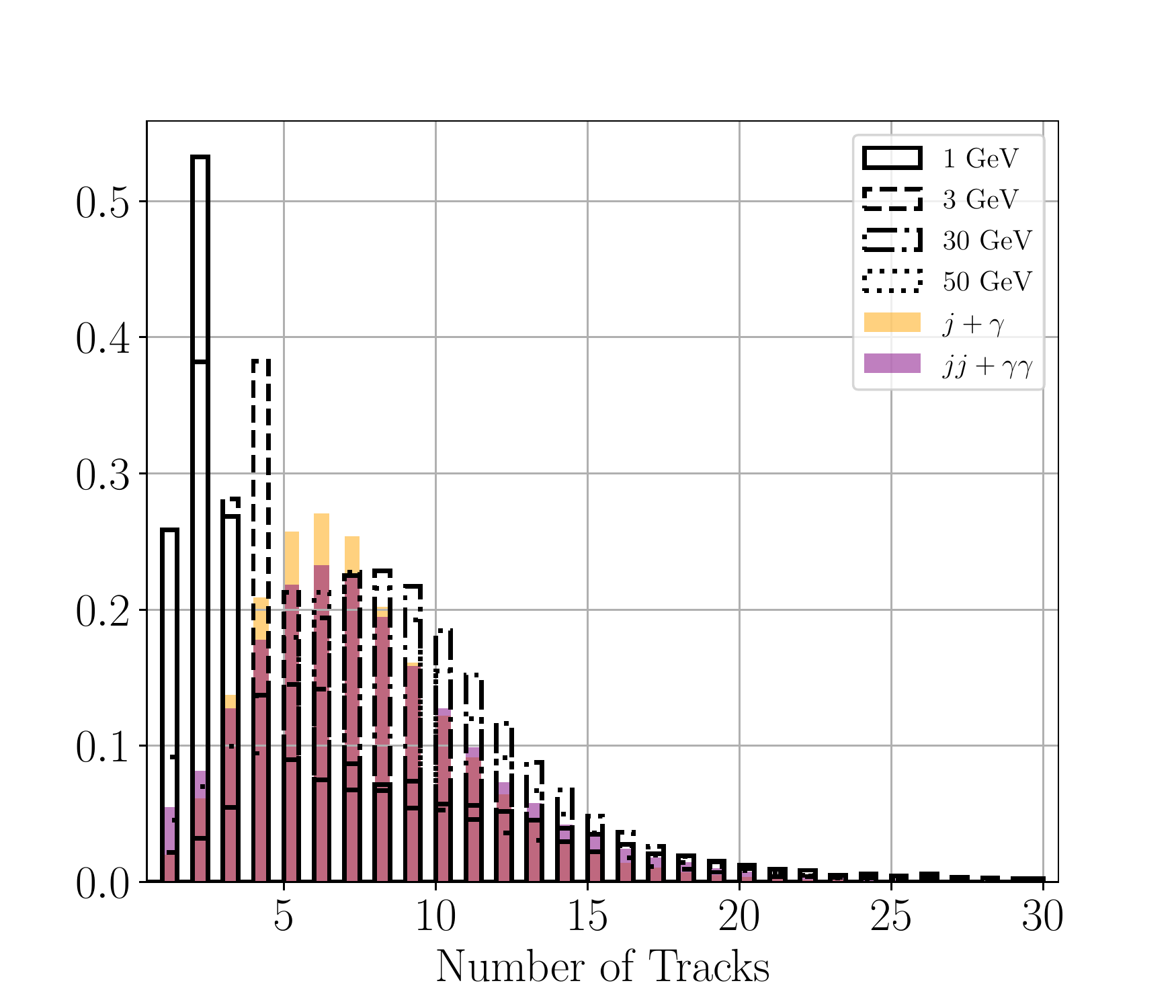}
\includegraphics[scale=0.45]{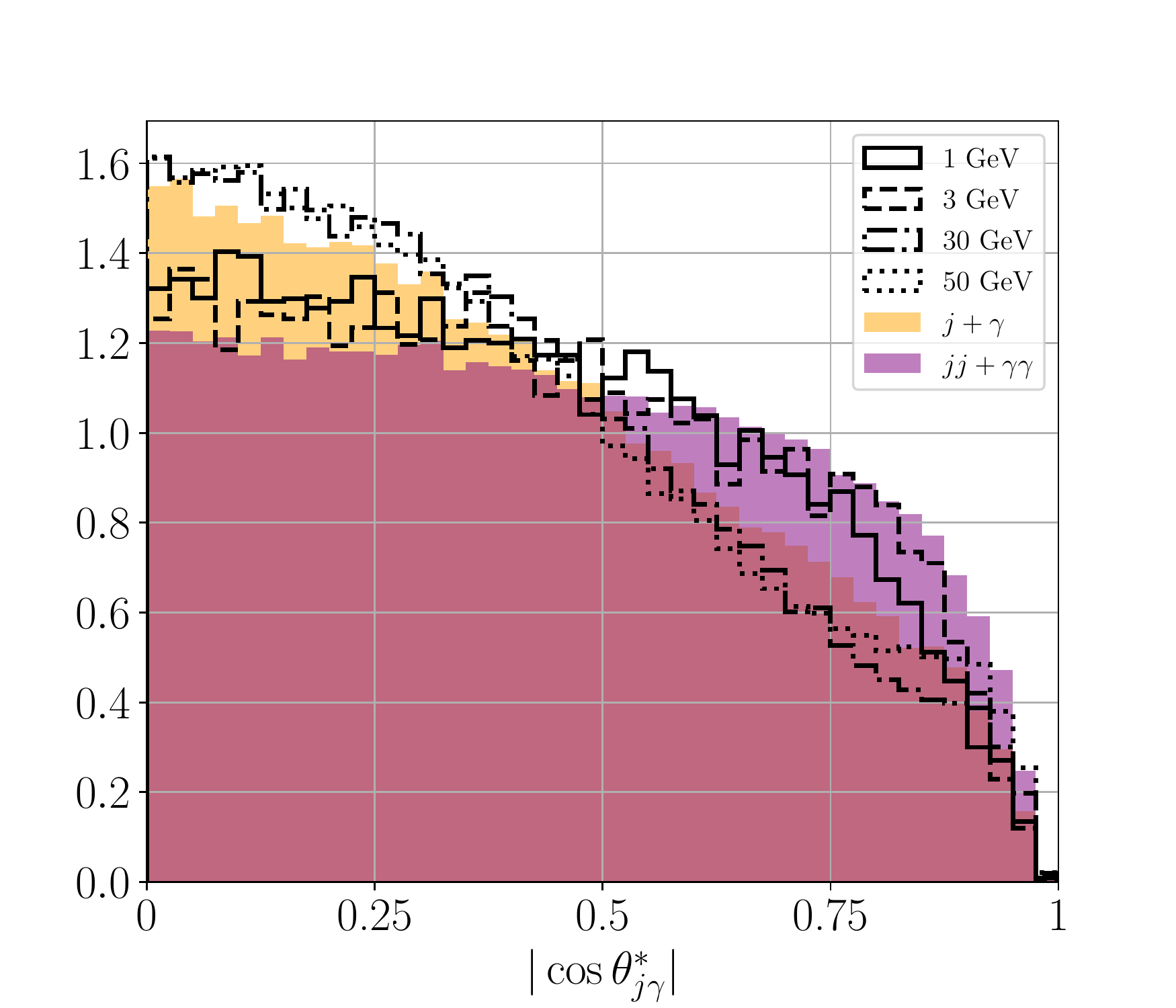}
\caption{The $\gamma(\gamma)$ transverse momentum (upper left), $j(j)$ transverse momentum (uuper right), the number of tracks of the leading jet (lower left), and $\cos\theta^*_{j\gamma}$ (lower right). We display the distributions for ALP masses of 1, 3, 30 and 50 GeV, and for the $j+\gamma$ and $jj+\gamma\gamma$ backgrounds. All distributions are normalized to unity.}
\label{fig:distr}
\end{figure}

Along with the invariant masses just described, we compound our events representation with the four additional distributions of Fig.~\eqref{fig:distr}. The upper row displays the transverse momentum of the $\gamma(\gamma)$ and $j(j)$ pair of the events at the left and right panels, respectively. In events where just one photon or jet is identified, the transverse momentum of the pair is substituted by the particle $p_T$. The lower panels show two other useful distributions, the number of tracks of the event at the left panel, and the $\cos\theta^*_{j\gamma}=\frac{1}{2}\tanh((\eta_j-\eta_\gamma)/2)$ at the right panel. This later variable effectively measures the cosine of the angle between the leading photon and the leading jet directions. For processes where the photon and the jet are colimated, $\theta^*$ tends to vanish. By its turn, signals from light ALP masses exhibit a smaller number of detectors tracks compared to backgrounds, whilst larger masses have slightly larger number of tracks.

The Higgs production cross section at the 13 TeV LHC is $48.5$ pb~\cite{Alwall:2014hca} in the gluon fusion process. The maximum Higgs to ALP branching ratio allowed by the LHC constraint of Eq.~\eqref{eq:h_total} is 32.2\% while the maximum branching fraction to $jj\gamma\gamma$ is $2\times(0.5)^2=0.5$ producing a maximum total branching ratio of around 16\%, that is it, a 7.8 pb signal at most. The dominant irreducible background, $j\gamma$, by its turn, has a cross section of $3.02\times 10^4$ pb including an extra jet contribution. The production cross section of the dominant reducible background of QCD jet pair production is huge, amounting to $2.1\times 10^8$ pb after the basic cuts of Eq.~\eqref{eq:basic_cuts}. The $jj\gamma\gamma$ cross section is $24.3$ pb after the basic cuts. As we discussed, the first step in reducing the backgrounds is by requiring the tight photon isolation criteria that suppresses the $jj$ background, otherwise, observing light ALP signals would be impossible. The isolation requirements affect most the light ALP signals, reaching a minimum efficiency of 5\% for $m_a=3$ GeV. On the other hand, they allow less than 1 in every $10^4$ events of jet pair production, approximately, bringing the $jj$ contamination to a fraction of the $j\gamma$ background. The photon isolation efficiency for $j\gamma$ and $jj\gamma\gamma$ backgrounds are 58\% and 77\%, respectively.
%

In order to automatically adapt the selection strategy to different ALP masses and optimize the signal significance, we used an optimization algorithm based on Tree Parzen Estimation (TPE) to search for the best cuts on the eight variables of Figs.~\eqref{fig:lines} and \eqref{fig:distr} with aid of \texttt{Hyperopt}~\cite{Bergstra:2013,Alves:2017ued}. Five thousand searches for each mass were performed seeking to maximize the statistical significance calculated from
\begin{equation}
    N_\sigma = \frac{S}{\sqrt{B+(\varepsilon_B B)^2}}, 
    \label{eq:signif}
\end{equation}
where $S(B)$ is the number of signal(total background) events for a given integrated luminosity, and $\varepsilon_B$ represents a systematic uncertainty in the total background rate. The systematic uncertainty were fixed at 0.01 to perform the searches in such a way that the algorithm learns to tame the systematics, raising the $S/B$ ratio~\cite{Alves:2017ued}. As an example, we show, in Eq.~\eqref{eq:cuts10}, the cuts that maximize the signal significance metric of Eq.~\eqref{eq:signif} for an ALP mass of 10 GeV
\begin{eqnarray}
&& 0.4 < M_J < 12.4\; \hbox{GeV},\; 0.9 < M_{jj} < 13.6\; \hbox{GeV},\; |M_{jj\gamma\gamma}-m_h| < 9.4\; \hbox{GeV}, \nonumber\\
&& 66 < p_T^{\gamma\gamma} < 237.5\; \hbox{GeV}, 37 < p_T^{jj} < 74\; \hbox{GeV},\nonumber \\
&& 14 < n_{tracks} < 33,\; 0 < |\cos\theta^*_{j\gamma}| < 0.6\; .
\label{eq:cuts10}
\end{eqnarray}

In this case, the algorithm learned that searching for a peak in the leading jet mass and $jj$ mass distributions is more efficient than in the $\gamma\gamma$ spectrum and, indeed, the windows were the cuts are placed in those distributions contain the mass value of 10 GeV. Moreover, note that the $p_T^{\gamma\gamma}$ and $p_T^{jj}$ cuts focus around $m_h/2$ where these distributions should be peak for the signals.  This cut strategy leads to a signal significance of $20.8\sigma$, after 3 ab$^{-1}$, at a point of parameters space  where $C_{ah}^{eff}=1\; \hbox{TeV}^{-2}$ and $BR(h\to aa)=21.7$\%. The overall efficiencies for the signal, $j+\gamma$, and $jj+\gamma\gamma$ backgrounds read $2\times 10^{-4}$, $1.4\times 10^{-7}$, and $2.8\times 10^{-6}$, respectively. Other masses present similar combined cut and identification efficiencies.
\begin{figure}[t!]
\includegraphics[scale=0.45]{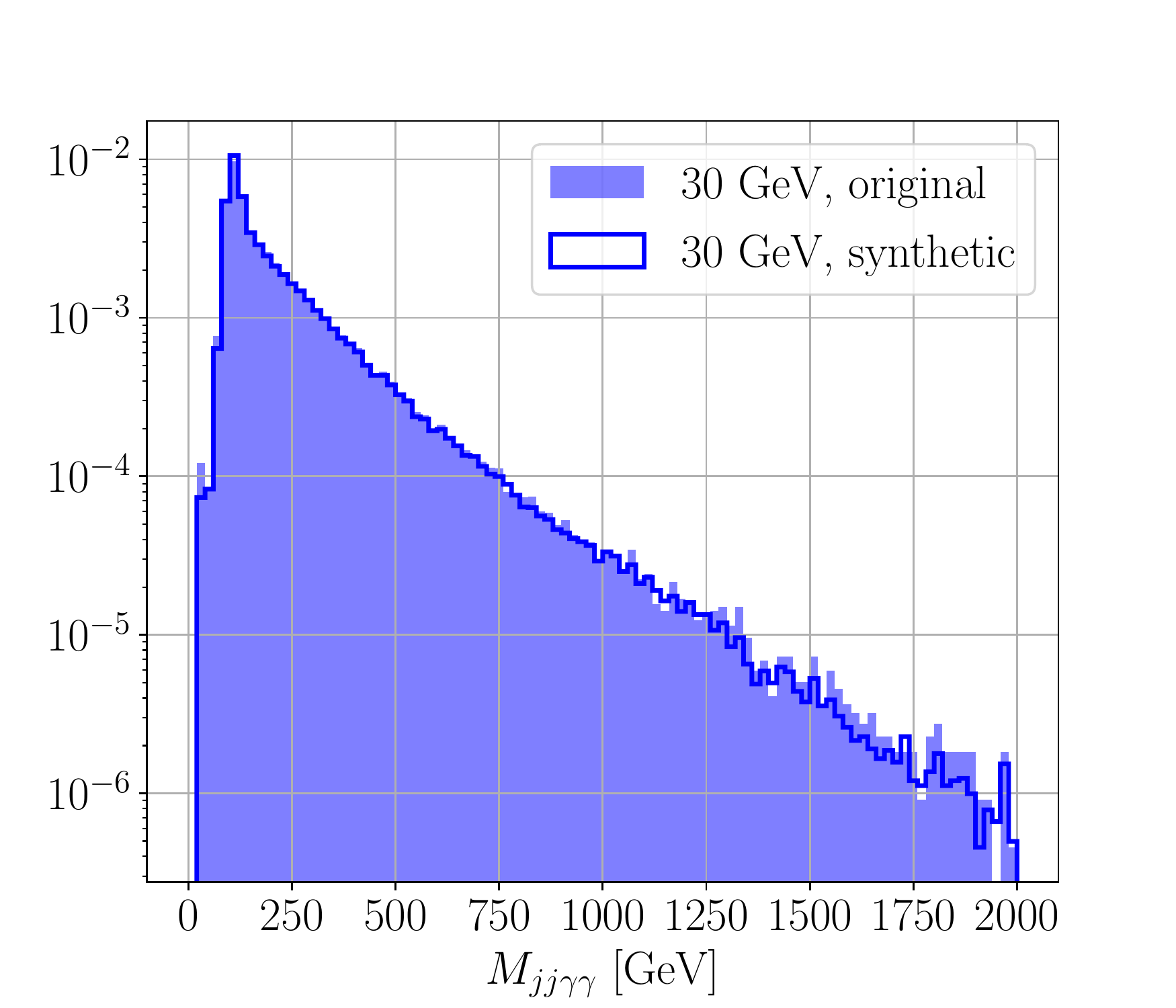}
\includegraphics[scale=0.45]{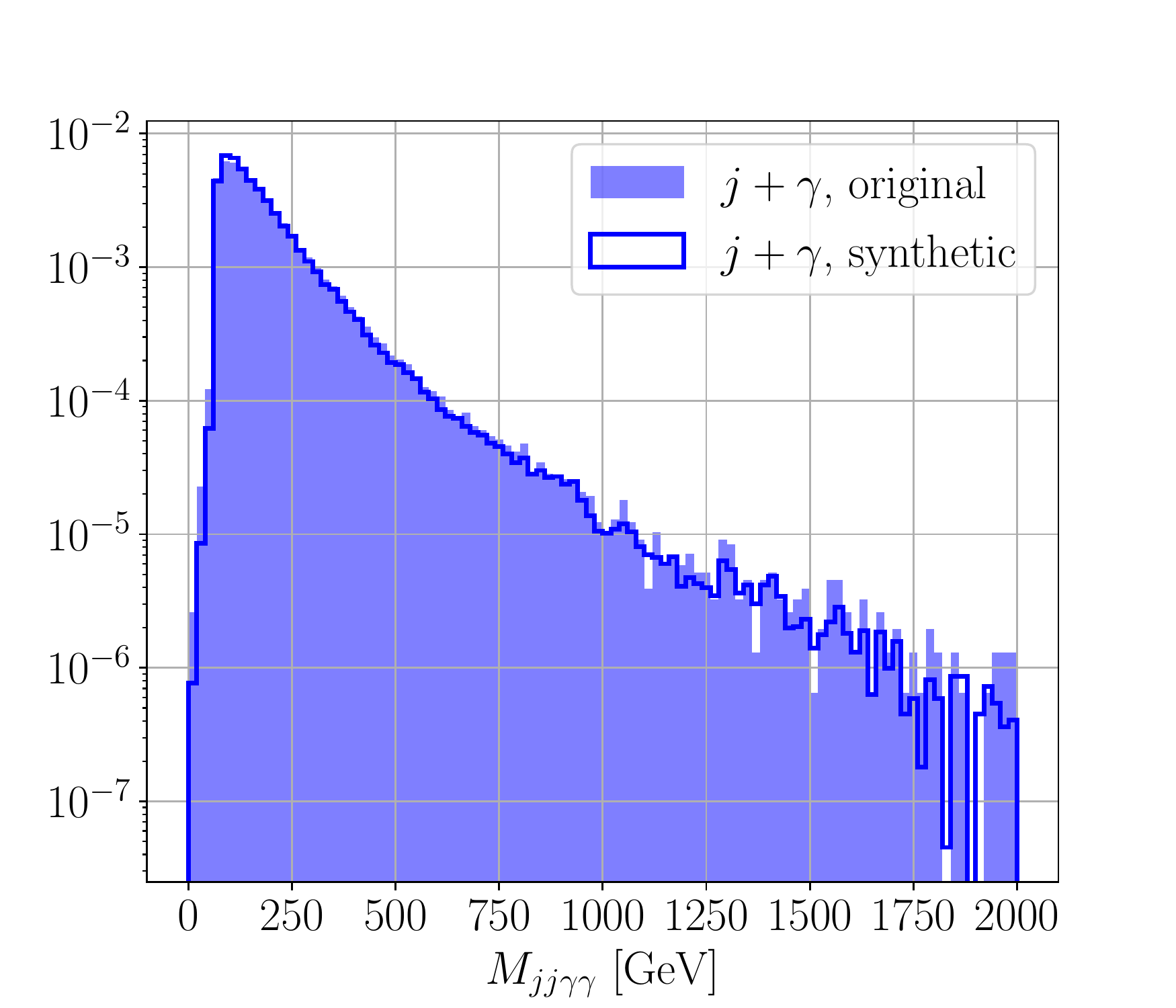}
\caption{Comparison between the original MC events generated by \texttt{MadGraph} (blue shaded histrograms) and the synthetic ones generated with the ADASYN method (solid blue). We display the normalized $M_{jj\gamma\gamma}$ distribution for the the ALP signal ($m_a=30$ GeV) at the left panel, and for the $j+\gamma$ background at the right panel.}
\label{fig:sintetic}
\end{figure}

Such small cut efficiencies pose a problem to the analysis once it is very time consuming to generate sufficiently Monte Carlo events to a reliable estimate. We overcome this difficulty by generating a large number of synthetic events using an well known data augmentation machine learning technique to tackle imbalanced datasets. In both machine learning studies and phenomenological analysis as ours, classifying events with good accuracy demands balanced datasets where the relative proportion of class instances are not too different from each other, otherwise the classification algorithms might  not be able to learn to identify instances of the minority class at the same time its accuracy performance is misleadingly high. One of those techniques is the Synthetic Minority Oversampling TEchnique~\cite{DBLP:journals/corr/abs-1106-1813}, or SMOTE, for short, and it is based on a very simple idea. First, an event is randomly chosen from the dataset and its $k$ nearest neighbours are found, then one of these neighbours is again randomly chosen. After determining those two events, a line is drawn connecting them in the representation space of the event (the space of kinematic and other variables of the tabulated event). Finally, a point in this line between those two events is randomly picked according to some distribution. In SMOTE, this distribution is uniform. In this work, we actually used the Adaptive Synthetic Sampling Method~\cite{4633969}, or ADASYN, that chooses a new point according to a local density distribution of points, thus trying to mimic, or adapt, to the true local density of points in the vicinity of the chosen event that seeds the generation of the synthetic point.~\footnote{Neural networks based data augmentation in high energy collision simulations have been addressed, for example, in Refs.~\cite{Butter:2020qhk,Butter:2020tvl,Chen:2020uds}. A detailed studied to compare the performance of our proposed $k$NN approach to other ones is due.}

In order to generate synthetic data, we firstly produced a few million MC samples from \texttt{MadGraph} for signals and backgrounds. After isolation, identification and basic cuts have been imposed, 10 to 100 thousand events survive depending on the process. These samples are then used to iteratively seed the generation of synthetic samples. To do so, we split the MC samples into two randomly chosen disjoint sets where the minority one contains a fraction $\alpha$ of the total number of samples. The algorithm then produces a number of synthetic samples to turn the size of two sets equal, that is it, $1-2\alpha$ times the size of the original MC event samples. We repeat this procedure until we get the desired number of samples. In Fig.~\eqref{fig:sintetic}, we show the $M_{jj\gamma\gamma}$ distribution for the signal (left panel) and the $j+\gamma$ (right panel) background. Just like this kinematic variable, we checked that synthetic and original samples generated by \texttt{MadGraph} display very similar shapes for all the other variables. Moreover, we also observed that the algorithm is able to populate sparse portions of the features space, effectively producing events in the tails of the distributions and helping to better estimate the cut efficiencies.

With this technique, we generated ${\cal O}(10^6)$ events to each background component and the signals. We also generated around $10^7$ $jj$ events to check its very small efficiency. These events were used to tune the cut strategy for several ALP masses from $0.5$ to $60$ GeV, fixing an 1\% systematic uncertainty in the total background rate as discussed earlier. Before we present the results of our analysis, let us investigate the points of the parameters space considered in this work that satisfy the constraints of Section~\ref{sec:constraints}.

\section{Scanning the parameters space}
\label{sec:models}
\begin{figure}[t!]
\includegraphics[scale=0.45]{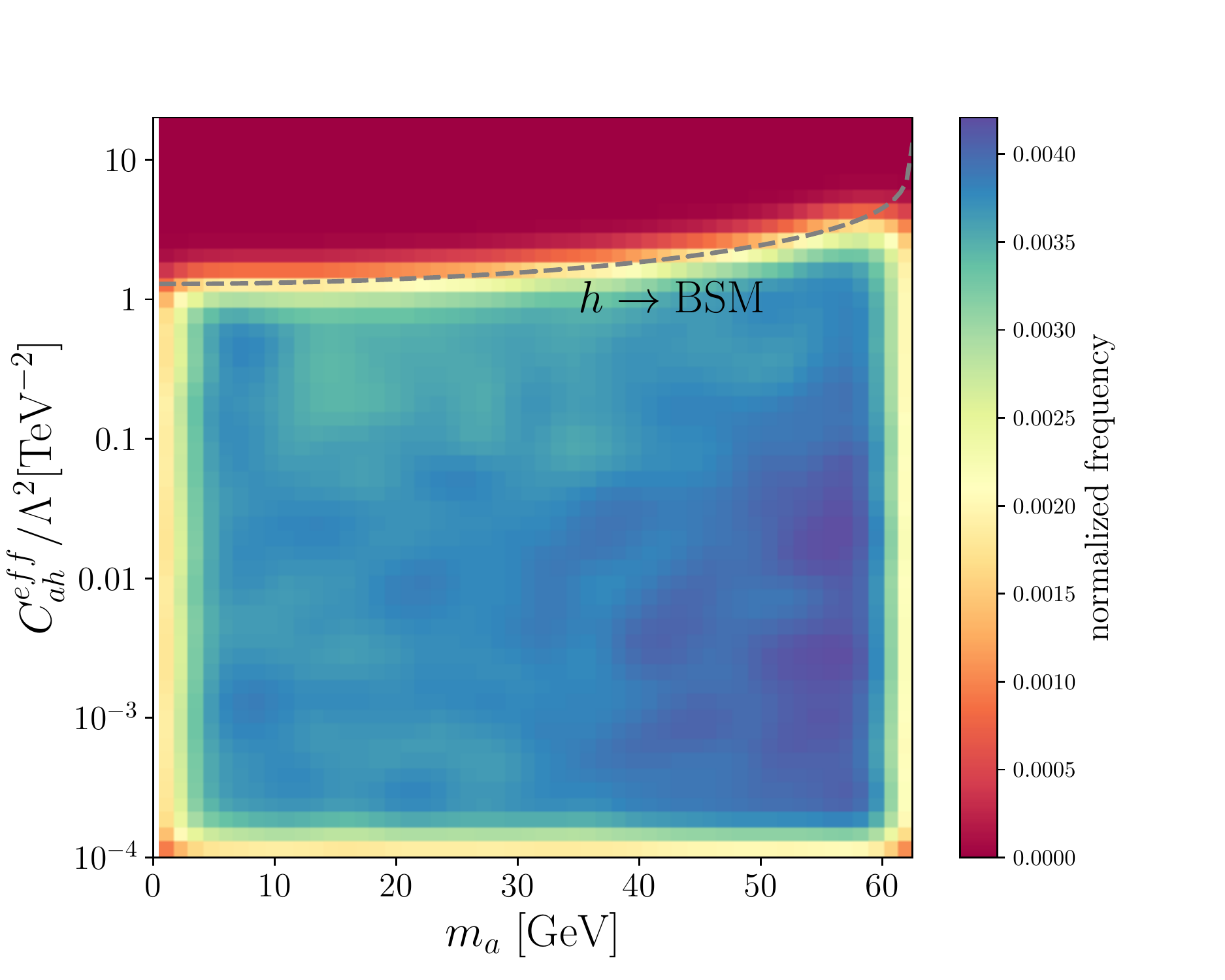}
\includegraphics[scale=0.45]{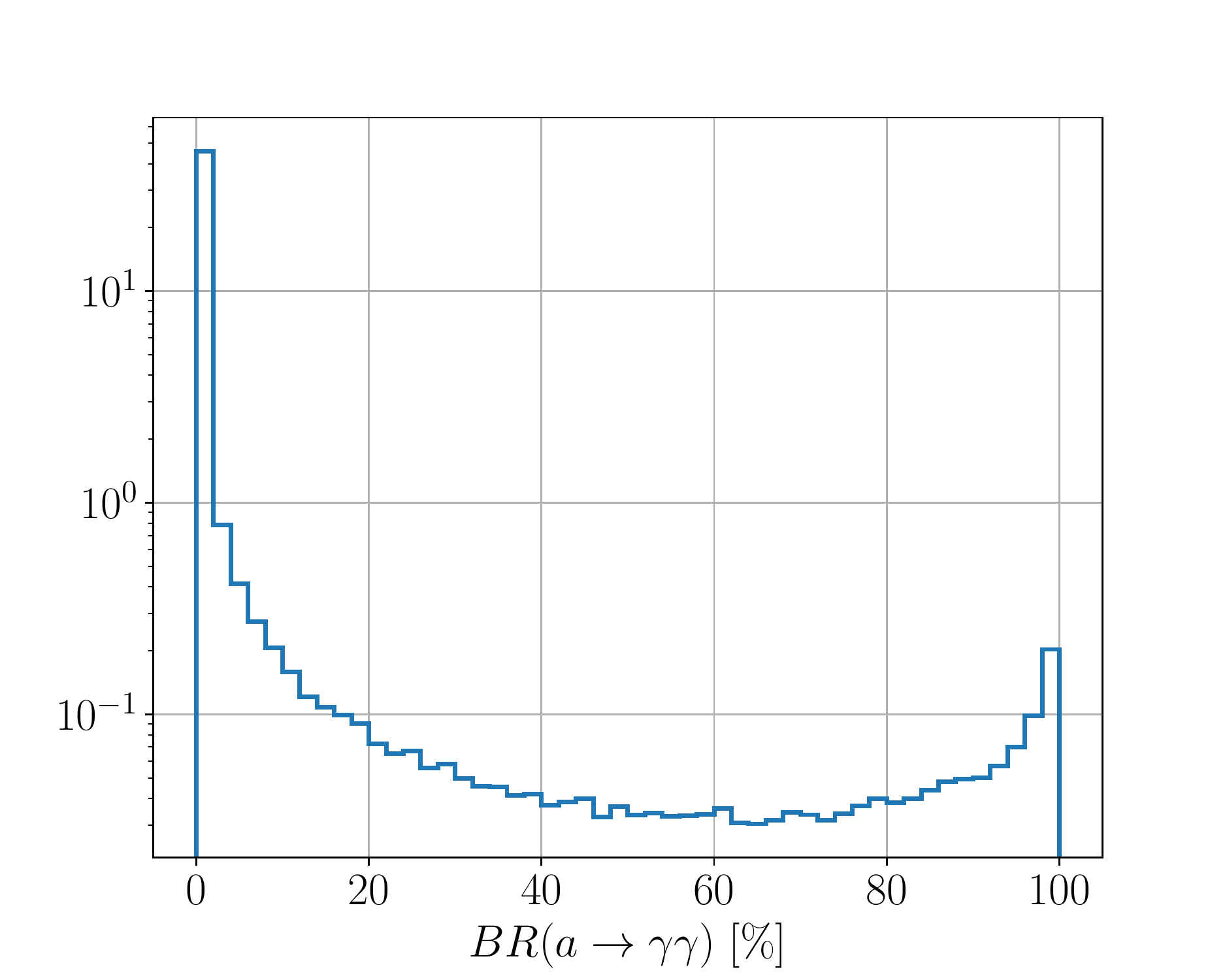}

\caption{At the left panel, we show the density of points of the parameters space Eq.~\eqref{eq:params} after imposing the collider constraints of Section~\ref{sec:constraints} in the  $C_{ah}^{eff}/\Lambda^2$ {\it versus} $m_a$ plane. Cooler regions have larger density of points. The dashed line represents the constraint from Eq.~\eqref{eq:h_total}. At the right panel, we plot the histogram of the normalized frequency of valid points for a given ALP branching ratio to photons.}
\label{fig:models}
\end{figure}

 Before proceeding to the results section, we need to estimate how many points of the parameters space of the model could eventually be probed at the LHC. For that goal, we scan a relevant portion of the parameters space defined by
 \begin{eqnarray}
 && \frac{k_{GG}}{\Lambda}\in [10^{-3},4\pi]\; \hbox{TeV}^{-1},\;\;  \frac{k_{\gamma\gamma}}{\Lambda}\in [10^{-3},4\pi]\; \hbox{TeV}^{-1},\;\; \frac{C_{ah}^{eff}}{\Lambda^2}\in [10^{-4},4\pi]\; \hbox{TeV}^{-2},\;\; \nonumber\\
&& m_a\in [0.5,62.5]\; \hbox{GeV}. 
\label{eq:params}
 \end{eqnarray}
 
 Despite we have chosen a different and more convenient basis to parametrize our model, some of the constraints of Section~\ref{sec:constraints} can only be translated to the original parameters contained in the Lagrangian of Eq.~\eqref{eq:Lagrangian2}. For this reason, we first draw samples from the original parameters of Eq.~\eqref{eq:params}, apply the constraints of Eqs.~\eqref{eq:ldecay}--\eqref{eq:Zphoton}, and then map those points that passed the requirements onto our chosen parameters space: $C_{ah}^{eff}/\Lambda^2$, $BR(a\to \gamma\gamma)$ and the ALP mass, $m_a$, as discussed in Section~\ref{sec:eft}.
 
 After the random draw of $2\times 10^6$ points from log-uniform distributions and imposing all the constraints from Eqs.~\eqref{eq:ldecay} to \eqref{eq:Zphoton}, around 20\% of the points passed all those constraints. Eq.~\eqref{eq:h_total}, limiting the largest branching ratio of the Higgs boson into new modes, and Eq.~\eqref{eq:Zphoton}, which constrains photonic decays of the $Z$ boson, are the harder ones.
 
 In Fig.~\eqref{fig:models}, left panel, we show the result of the scan displaying the density of the parameters effectively used to parametrize the model for points which evade the constraints imposed in the $C_{ah}^{eff}/\Lambda^2$ {\it versus} $m_a$ plane, the parameters which are interested to display our results. We see that Eq.~\eqref{eq:h_total} prohibits points above the dashed line, but weakens as $m_a$ approaches $m_h/2$. As a result, an slightly larger concentration of points occurs for heavier masses near $m_h/2$, yet we have a more or less uniform density of points in the whole ALP mass range.
  
  At the right panel of  Fig.~\eqref{fig:models}, we display the density of points in terms of the branching ratio of the ALP into photons. The majority of points are found for very small values of the branching ratios, but we can also find points where $BR(a\to\gamma\gamma)\approx 50$\%. From Eq.~\eqref{eq:widths}, the branching ratio of the ALP into photons can be expressed as $1/(1+8k_{GG}^2/k_{\gamma\gamma}^2)$ and the ratio $k_{GG}^2/k_{\gamma\gamma}^2$ can assume values from $10^{-4}$ up to $10^4$ when parameters vary as in Eq.~\eqref{eq:params}. This ratio is evenly distributed around unit, giving the branching ratio distribution an almost symmetric shape. The constraints from Eqs.~\eqref{eq:hjjaa}, \eqref{eq:4gamas_high} and \eqref{eq:Zphoton}, however, eliminate those points with large $BR(a\to\gamma\gamma)$, and only those ones where the Higgs-ALP coupling is too small and/or $k_{BB}\approx k_{WW} \gg k_{GG}$ can have a large ALP to photons decay rate. Now, we present the results of our analysis.

\begin{figure}[t!]
\includegraphics[scale=0.45]{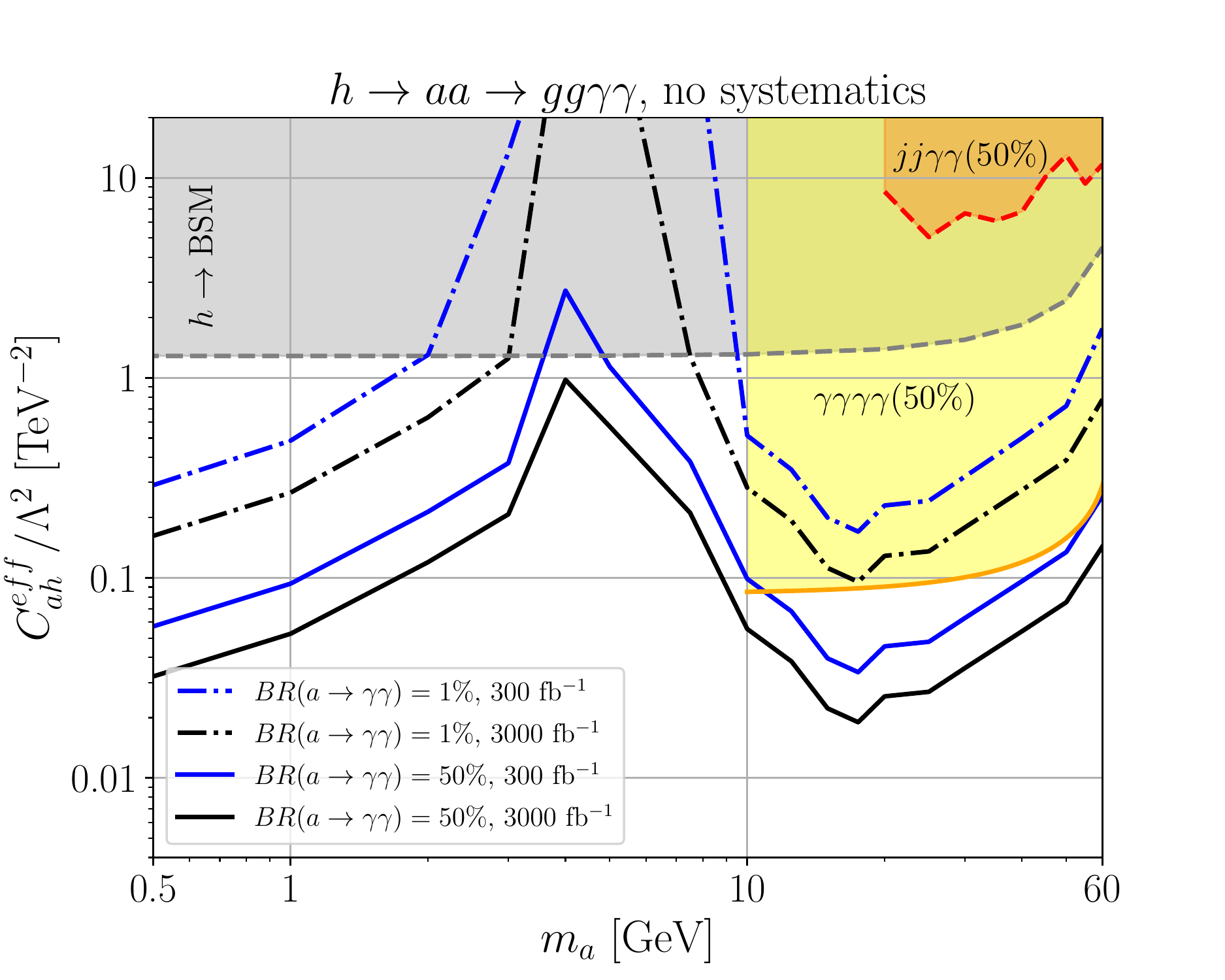}
\includegraphics[scale=0.45]{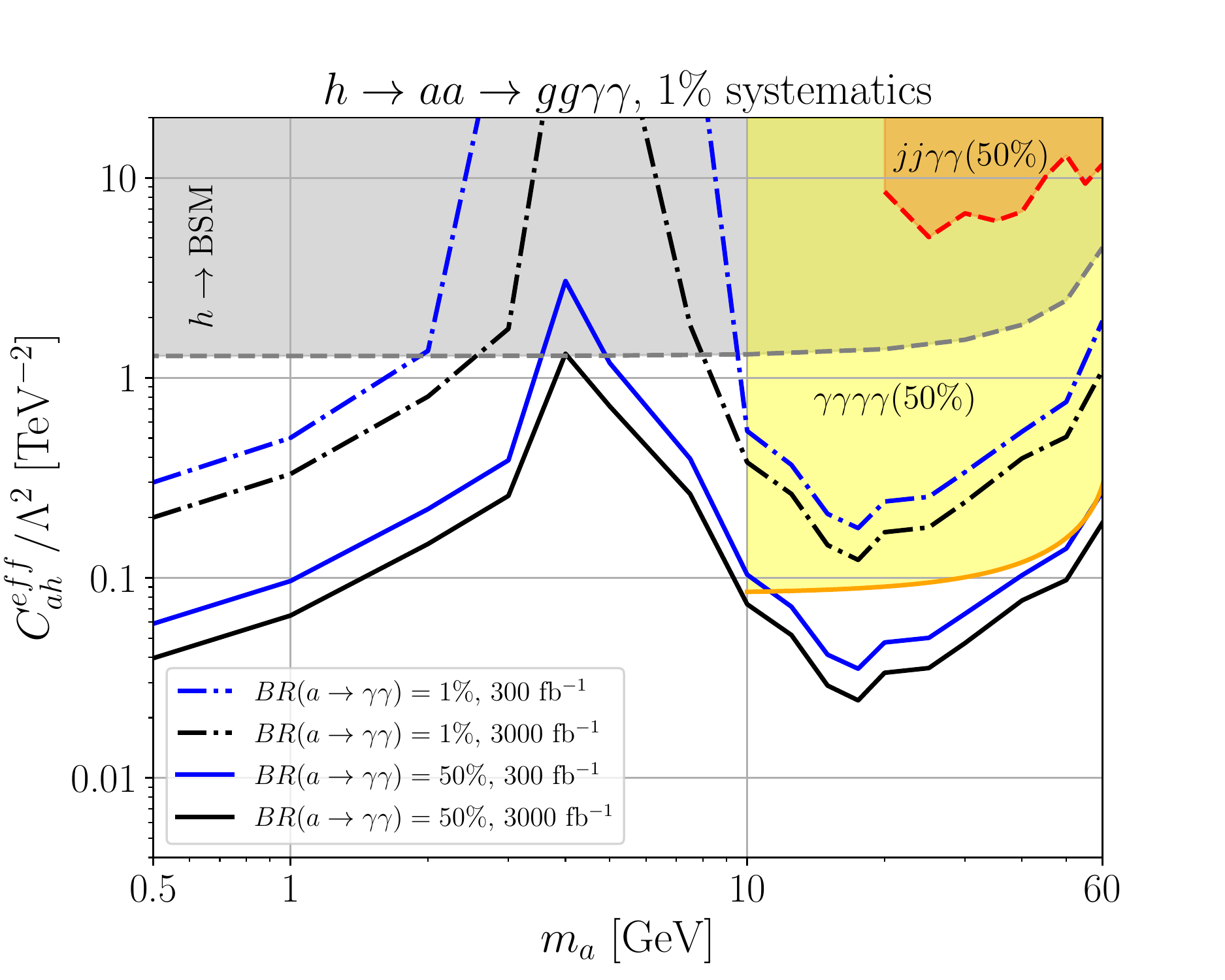}\\
\includegraphics[scale=0.45]{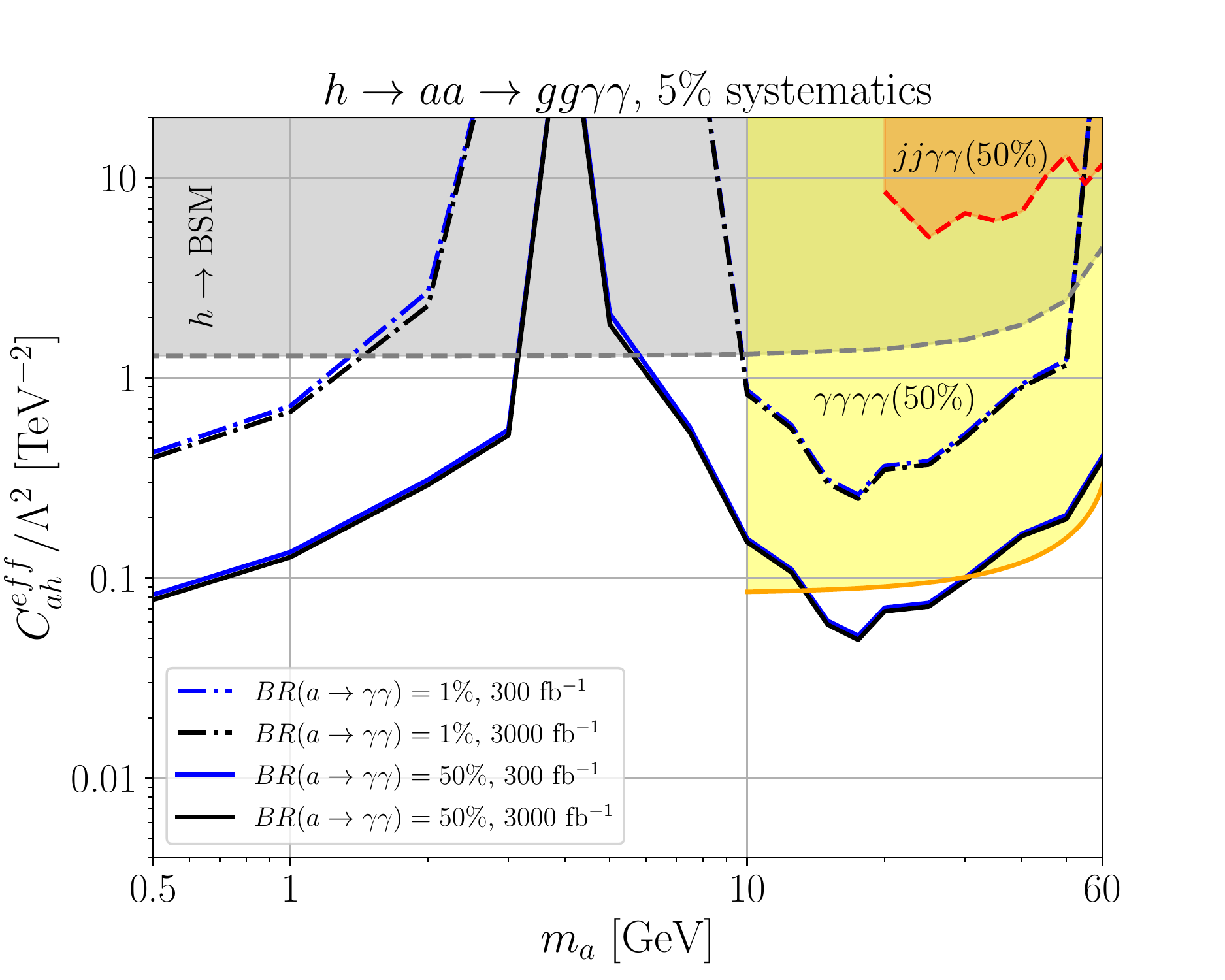}
\includegraphics[scale=0.45]{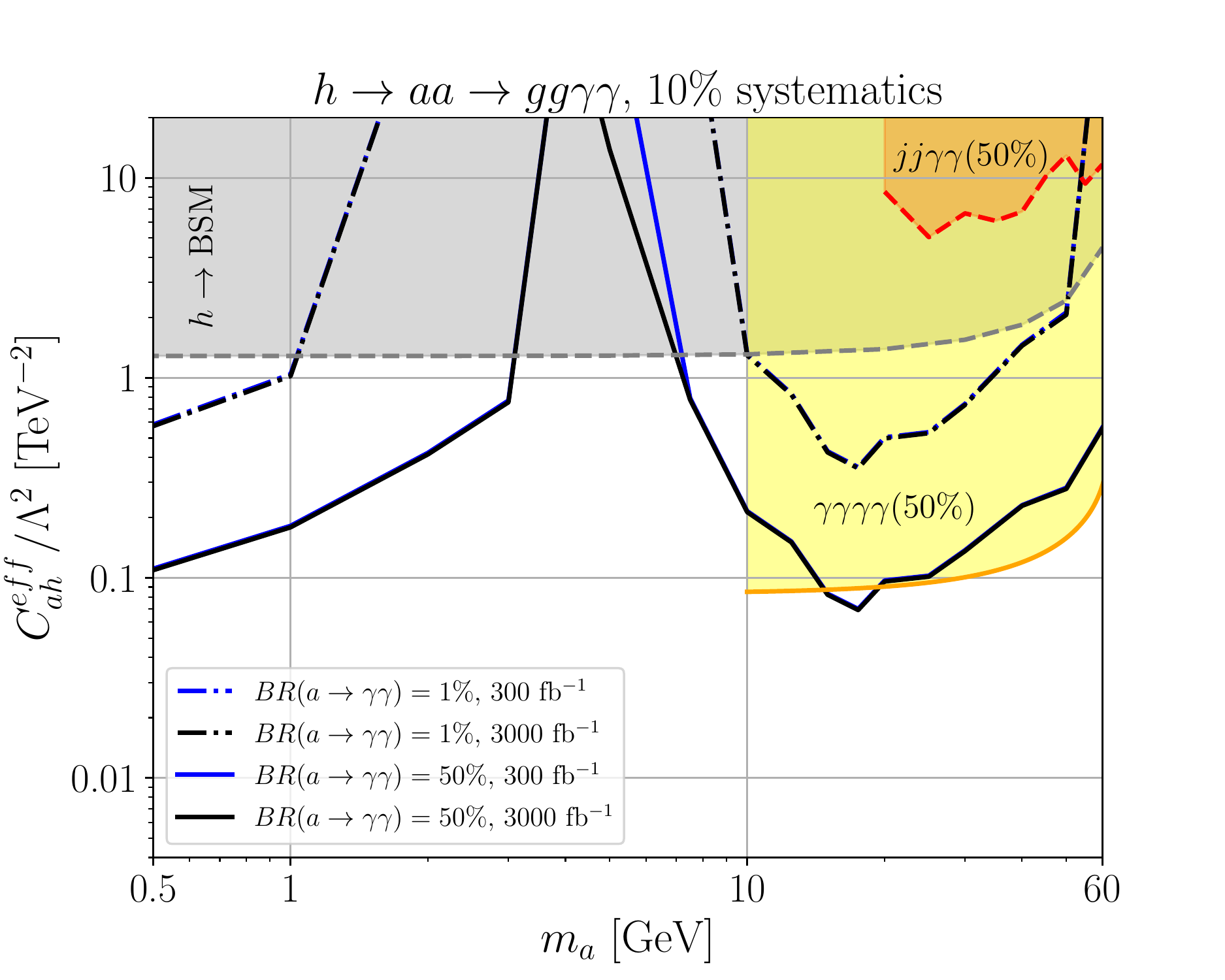}
\caption{The 95\% CL limits on the effective ALP-Higgs coupling (in TeV$^{-2}$) {\it versus} the ALP mass (in GeV). The orange shaded area denotes the limits from Eq.~\eqref{eq:hjjaa}, while the yellow shaded area represents the limits from Eq.~\eqref{eq:4gamas_high}, assuming $BR(a\to\gamma\gamma)=50$\% in both cases. The blue(black) lines represent the limits obtained in the $j(j)+\gamma(\gamma)$ channel assuming 300(3000) fb$^{-1}$ of data, while the solid(dot-dashed) curves represent the case where branching ratio of the ALP to photons is 50\%(1\%). We show four different systematics scenarios: $\varepsilon_B=0,1,5,10$\%. The gray shaded region ($h\to \hbox{BSM}$) is excluded, at 95\% CL, by the LHC constraint on the Higgs total width, Eq.~\eqref{eq:h_total}.}
\label{fig:res}
\end{figure}
\section{Results}
\label{sec:results}

 After tuning the cuts to get the maximum sensitivity for each ALP mass, we take the cut efficiencies to estimate the signal significance by computing the branching ratios at a given point of the parameters space. The cut efficiency depends only on the ALP masses, which modify the kinematic distributions. Our goal is to evaluate the sensitivity of the $j(j)+\gamma(\gamma)$ channel and compare it against the $jj\gamma\gamma$ and $\gamma\gamma\gamma\gamma$ channels given by Eqs.~\eqref{eq:hjjaa} and \eqref{eq:4gamas_high}, respectively, in the $C_{ah}^{eff}/\Lambda^2$ {\it versus} $m_a$ plane. 

 We show in Fig.~\eqref{fig:res}, the regions of the $C_{ah}^{eff}/\Lambda^2$ {\it versus} $m_a$ plane that can be probed at 95\% CL in the $j(j)+\gamma(\gamma)$ channel in four different systematics scenarios: statistics domination ($\varepsilon_B=0$) and 1\% systematics at the left and right upper panels, respectively, and 5\% and 10\% at the left and right lower panels, respectively. Dot-dashed lines in that figure represent points of the parameters space for which $BR(a\to\gamma\gamma)=1$\%, while the solid curves represent the case where $BR(a\to\gamma\gamma)=50$\%.  We found that systematic uncertainties dominate the errors for $\varepsilon_B \gtrsim 3$\%. As a consequence, if the background rates can be precisely determined, we should expect to observe improvements of  statistical limits with increasing luminosity, otherwise, the limits for 300 and 3000 fb$^{-1}$ will look very similar as we see in the lower panels of Fig.~\eqref{fig:res}. 

As expected, the regions of the parameters space that can be probed for a fixed luminosity improve as $BR(a\to\gamma\gamma)$ approaches 50\%. The maximum sensitivity occurs for ALP masses close to 0.5 and 20 GeV, two well distinct regimes where the majority of events contain $j+\gamma$ pairs and $jj+\gamma\gamma$ pairs, respectively, reaching $C_{ah}^{eff}/\Lambda^2 \approx 0.02$ TeV$^{-2}$. Even for the high systematics scenario of 10\%, an effective ALP-Higgs coupling of order $0.1$ TeV$^{-2}$ can be excluded at the 95\% CL. On the other hand, unless the systematics are well controlled, a mass gap between 2 and 5 GeV, approximately, might still remain unreachable at the 13 TeV LHC. Yet, the very light end of the former mass gap region, between 0.5 and 2 GeV, can be probed in the $j(j)+\gamma(\gamma)$ channel, contrary to the four photons and the fully resolved $jj\gamma\gamma$ channels. 

In the mass region where $m_a > 10$ GeV, we display the current constraint from the ATLAS search for $h\to\gamma\gamma\gamma\gamma$, Eq.~\eqref{eq:4gamas_high}. This search was performed at the 8 TeV LHC after 20.3 fb$^{-1}$, so we expect that future reevaluations of these limits get significantly stronger. These results, however, are very sensitive to $BR(a\to\gamma\gamma)$. In Fig.~\eqref{fig:res}, we display the region of the parameters excluded at 95\% CL by this search channel if $BR(a\to\gamma\gamma)=50$\% (solid yellow line). The 1\% case does not constraint the region of the parameters displayed in the plot. What we basically learn is that $j(j)+\gamma(\gamma)$ is competitive if ALP branching ratio into photons is small disregard the level of systematics affecting the jets plus photons channel. On the other hand, $h\to\gamma\gamma\gamma\gamma$ will remain as the best option to constraint models where ALP branching ratio into photons is large if $m_a>10$ GeV. As we anticipated, the current limits from Eq.~\eqref{eq:hjjaa} do not harm the particular effective model that we are considering in this work. This experimental search involved 36.7 fb$^{-1}$ of 13 TeV LHC data~\cite{Sirunyan:2018ouh} and it is also expected to improve in the future. 

Comparing the Fig.~\eqref{fig:res} against the left panel of Fig.~\eqref{fig:models}, we see that a large number of points that evade the collider constraints can be probed in the $j(j)+\gamma(\gamma)$ channel in the whole region between 0.5 and 60 GeV. From the right panel of Fig.~\eqref{fig:models}, we learned that most points have $BR(a\to\gamma\gamma)$ close to 1\%, and that a branching ratio close to 50\%, for which the sensitivity is maximum, is rarer. Nevertheless, we found that small branching ratios for $a\to\gamma\gamma$ can be probed even with 300 fb$^{-1}$ if the systematic uncertainties could be kept under control.

\begin{figure}[t!]
\includegraphics[scale=0.45]{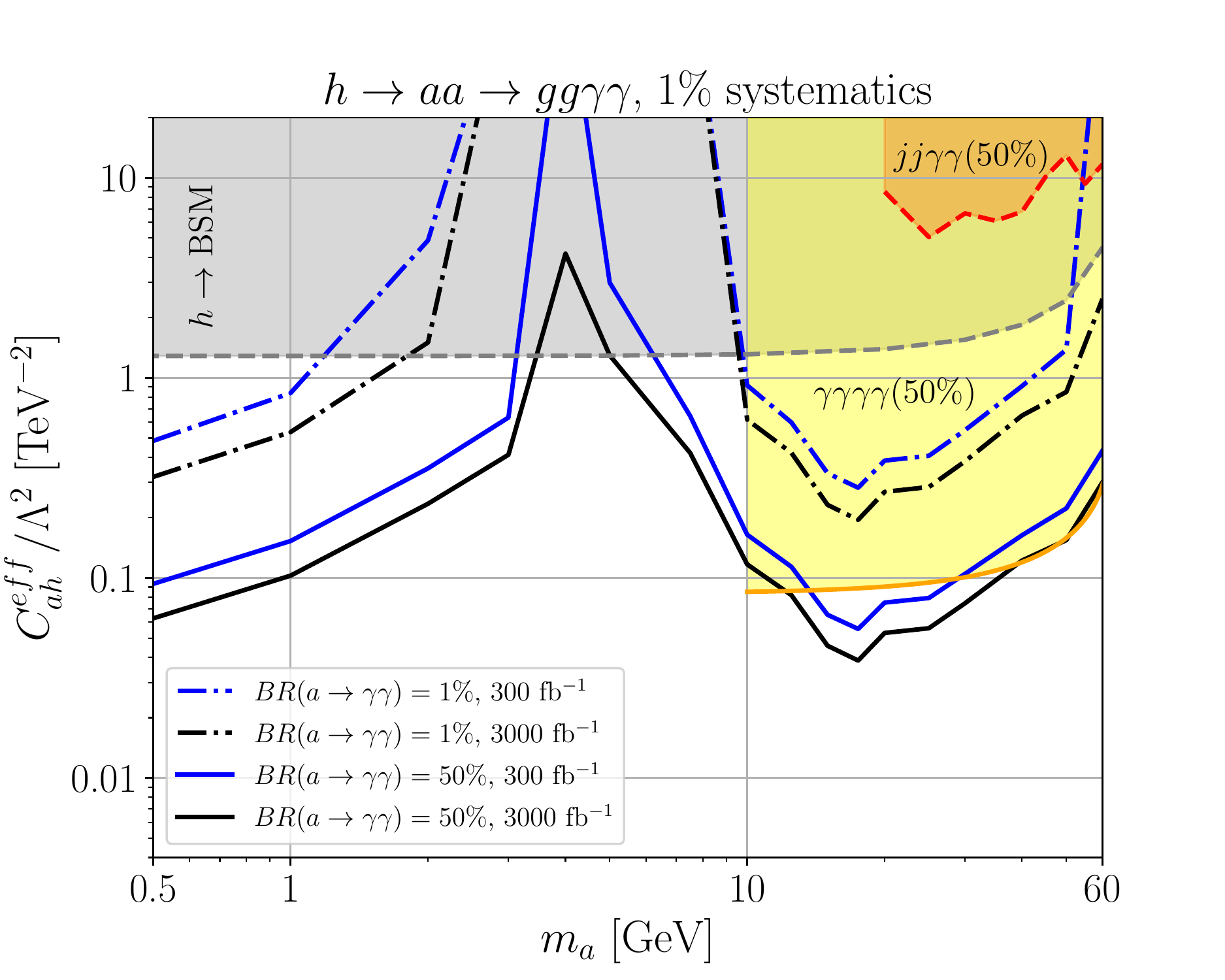}
\includegraphics[scale=0.45]{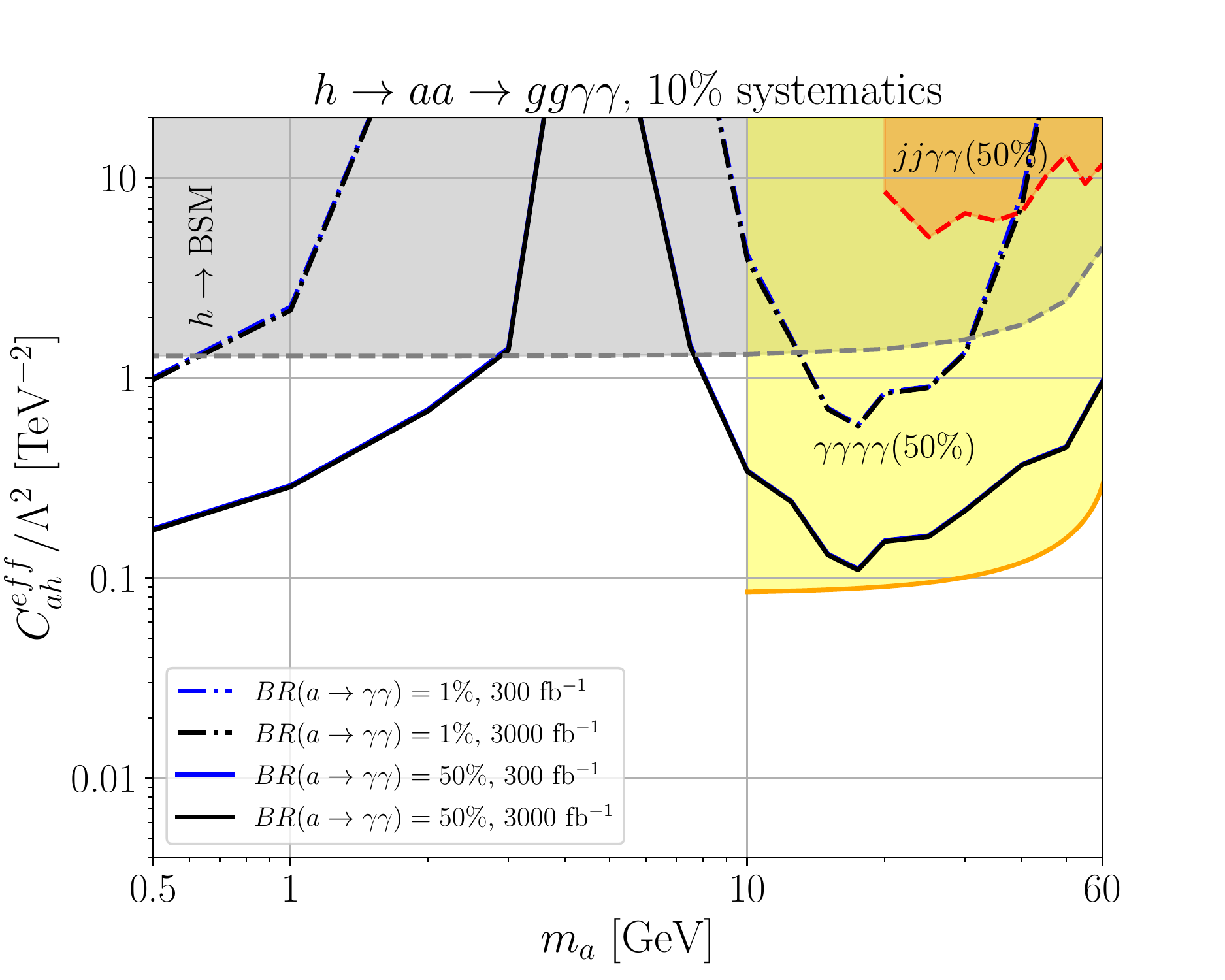}
\caption{The region (above the lines) in the effective ALP-Higgs coupling (in TeV$^{-2}$) {\it versus} the ALP mass (in GeV) where a discovery ($5\sigma$) of $h\to aa\to j(j)+\gamma(\gamma)$ is possible.  The yellow shaded area represents the excluded region from Eq.~\eqref{eq:4gamas_high} and orange shaded one, the limits from Eq.~\eqref{eq:hjjaa}, assuming $BR(a\to\gamma\gamma)=50$\% in both cases. The blue(black) lines bound the $5\sigma$ regions obtained in the $j(j)+\gamma(\gamma)$ channel assuming 300(3000) fb$^{-1}$ of data, while the solid(dot-dashed) curves represent the case where branching ratio of the ALP to photons is 50\%(1\%). We show two different systematics scenarios: $\varepsilon_B=1,10$\%. The gray shaded region ($h\to\hbox{BSM}$) is excluded, at 95\% CL, by the LHC constraint on the Higgs total width, Eq.~\eqref{eq:h_total}.}
\label{fig:discover}
\end{figure}

In Fig.~\eqref{fig:discover}, we show the regions of the  $C_{ah}^{eff}/\Lambda^2$ {\it versus} $m_a$ where $N_\sigma \geq 5\sigma$ for two systematics scenarios, 1\% and 10\% at the left and right panels, respectively. Assuming $BR(a\to\gamma\gamma)=1$\%, we see that little room is left for discovery that has not been excluded by the collider constraints in the low ALP mass region, however, there is a sizeable portion of the parameters space not excluded if $m_a>10$ GeV. The picture changes for $BR(a\to\gamma\gamma)=50$\%, now a discovery will be possible more probably in the mass region of $m_a<10$ GeV, even for 10\% systematics.

\section{Conclusions and prospects}
\label{sec:conclusion}

 If axion-like particles do exist, they should interact with the SM Higgs boson in an extended scalar sector. The Higgs-ALP coupling can be directly probed in decays of the Higgs into ALP pairs. Many possibilities exist to explore the yields of the ALP decays in different decay channels. An almost minimal approach is to assume that the new scalar couples at least with gluons and photons, besides the Higgs bosons, and parametrizing the signals with fewer parameters, namely, the branching ratio of ALPs to photons, the Higgs-ALP coupling and the ALP mass. The SM can be extended in many ways to account for new scalars with the proprieties of an ALP, so an effective field theory model is suitable for a general parametrization of those possibilities assuming that, apart from the ALP, all the new particles are heavy enough to be integrated out, validating the effective field theory approach.
 
 At colliders, the Higgs-ALP coupling has been searched for in processes with final state photons or leptons, mainly. These searches, however, loose sensitivity as soon as the ALP decays to hadrons are allowed, unless a large branching ratio to photons or leptons is assumed. In this work, we showed that the channel $pp\to h\to aa\to j(j)+\gamma(\gamma)$ might be able to probe Higgs-ALP couplings in regions of the parameters space where the photons channels cannot, especially in the mass region of 0.5 to 10 GeV. Our analysis automatizes the cut strategy by searching for Higgs and ALP resonances in various jets and photons masses combinations in resemblance to spectroscopy characterization of a material composition. This machine learning of the kinematic cuts to select signals enables us to place limits and estimate discovery prospects across the whole ALP mass region from 0.5 GeV, right after the hadronic decys are open, until 60 GeV, the Higgs to ALPs decay threshold.
 
 We found that if the systematic uncertainties in the background rates can be controlled, the whole ALP mass region from 0.5 to 60 GeV can be probed at 95\% CL even assuming an 1\% ALP to photons branching ratio after 300 fb$^{-1}$ of data. With 3000 fb$^{-1}$ and $BR(a\to\gamma\gamma)=50$\%, effective Higgs-ALP couplings down to 0.02 TeV$^{-2}$ can be excluded at 95\% CL for ALP masses close to 0.5 and 20 GeV but the $h\to aa\to \gamma\gamma\gamma\gamma$ might be a better option to probe the 10 to 60 GeV mass range. A scan of the parameters space of the model showed that the majority of the points that evade collider constraints can actually be probed in this proposed channel.  Discovery is possible in scenarios where the systematic uncertainties are under control and ALP branching ratio into photons is sizeable in the mass range of 0.5 to 10 GeV, except for the 2--5 GeV region. Overall, the $j(j)+\gamma(\gamma)$ channel is a better option than $h\to aa\to \gamma\gamma\gamma\gamma$ if the branching ratio of the ALP to photons is small and/or if the ALP is lighter than 10 GeV. 
 
 An important ingredient of our analysis is the suppression of the reducible $jj$ background which produces isolated photons from neutral mesons decays. For that goal, a tight photon isolation requirement is necessary, but at the cost of also suppressing the signals for ALP masses from 2 to 5 GeV, a mass region where the photons and jets from the colimated ALP decays start to get resolved by the detectors. For masses larger than $\sim 10$ GeV, the number of events with more than one isolated jet and photon increase and we can search for the $h\to aa$ with a more inclusive topology. If the branching ratio of ALP to photons is large enough, the $jj+\gamma\gamma$ might provide tighter constraints than the four photons channel in the 10 to 60 GeV region. Scenarios with masses lighter than 10 GeV can be better probed when the ALP decays are colimated and lead to a $j+\gamma$ pair, what happens more efficiently in the 0.5 to 2 GeV region. Overall, if systematic uncertainties could be tamed, the Higgs-ALP coupling can be probed at the HL-LHC in the whole mass region from 0.5 t0 60 GeV.
 
 A way around of not being forced to adopt very tight photon isolation criteria is looking for alternative Higgs production modes. For example, in associated $Zh$ and $Wh$ production~\cite{Martin:1997ns}, additional charged leptons and/or missing energy would hardly be mimicked by QCD $jj$ events, despite other sources of backgrounds would come into play. 
 The vector boson fusion channel also shows good perspectives~\cite{Aaboud:2018gmx} and could benefit from a smart cut selection strategy. Finally, in this work, we did not attempt to tune the photon and jet isolation requirements once it is too computationally expensive, but at the analysis level, relaxing or tightening those criteria  as the ALP mass varies would probably help to enhance the LHC sensitivity to Higgs-ALP interactions.

\section*{Acknowledgements}
The works of A. A. and A. G. D. are supported by the Conselho Nacional de Desenvolvimento Cient\'{\i}fico e Tecnol\'ogico (CNPq), under the grants 307265/2017-0 and 306636/2016-6, respectively. D. D. L. acknowledges the financial support from the Coordena\c{c}\~ao de Aperfei\c{c}oamento de Pessoal de N\'{\i}vel Superior - Brasil (CAPES) - 23 Finance Code 001.


\bibliographystyle{apsrev4-1} 
\bibliography{referencias}


\end{document}